\documentclass[sn-mathphys,Numbered]{sn-jnl}

\usepackage{graphicx}%
\usepackage{multirow}%
\usepackage{amsmath,amssymb,amsfonts}%
\usepackage{amsthm}%
\usepackage{mathrsfs}%
\usepackage[title]{appendix}%
\usepackage{xcolor}%
\usepackage{textcomp}%
\usepackage{manyfoot}%
\usepackage{booktabs}%
\usepackage{algorithm}%
\usepackage{algorithmicx}%
\usepackage{algpseudocode}%
\usepackage{listings}%
\usepackage{subcaption}
\usepackage{colortbl}
\usepackage{makecell}
\usepackage{verbatim}
\usepackage{verbatimbox}
\usepackage{fancyvrb}

\theoremstyle{thmstyleone}%
\theoremstyle{thmstyletwo}%
\newtheorem{example}{Example}%

\theoremstyle{thmstylethree}%

\newcommand{\node}[1]{\texttt{#1}}
\newcommand{\bneg}{\texttt{!}}
\newcommand{\band}{~\texttt{\&}~}
\newcommand{\bor}{~\texttt{|}~}
\newcommand{\floor}[1]{\lfloor #1 \rfloor}

\newenvironment{bnexample}{\begin{minipage}{0.96\textwidth}\centering\upshape\medskip}{\bigskip\end{minipage}}

\begin{document}

\title[Article Title]{An open problem: Why are motif-avoidant attractors so rare in asynchronous Boolean networks?}


\author[1,2,*]{\fnm{Samuel} \sur{Pastva}}\email{xpastva@fi.muni.cz}

\author[3]{\fnm{Kyu Hyong} \sur{Park}}\email{kjp5774@psu.edu}

\author[1]{\fnm{Ond\v{r}ej} \sur{Huvar}}\email{xhuvar@fi.muni.cz}

\author[4,5,*]{\fnm{Jordan C.} \sur{Rozum}}\email{jrozum@binghamton.edu}

\author[3,*]{\fnm{R\'eka} \sur{Albert}}\email{rza1@psu.edu}

\affil[1]{\orgdiv{Faculty of Informatics}, \orgname{Masaryk University}, \orgaddress{\street{Botanicka 68a}, \city{Brno},~\postcode{60200}, \country{Czech Republic}}}

\affil[2]{\orgname{Institute of Science and Technology Austria}, \orgaddress{\street{Am Campus 1}, \city{Klosterneuburg},~\postcode{3400}, \country{Austria}}}

\affil[3]{\orgdiv{Department of Physics}, \orgname{Pennsylvania State University}, \orgaddress{\street{Davey Laboratory}, \city{University Park}, \postcode{16802}, \state{Pennsylvania}, \country{USA}}}

\affil[4]{\orgdiv{Department of Systems Science and Industrial Engineering}, \orgname{Binghamton University (State University of New York)}, \orgaddress{\street{Engineering Building}, \city{Vestal}, \postcode{13850}, \state{New York}, \country{USA}}}

\affil[*]{Corresponding autor}



\abstract{Asynchronous Boolean networks are a type of discrete dynamical system in which each variable can take one of two states, and a single variable state is updated in each time step according to pre-selected rules. Boolean networks are popular in systems biology due to their ability to model long-term biological phenotypes within a qualitative, predictive framework. Boolean networks model phenotypes as attractors, which are closely linked to minimal trap spaces (inescapable hypercubes in the system's state space). In biological applications, attractors and minimal trap spaces are typically in one-to-one correspondence. However, this correspondence is not guaranteed: motif-avoidant attractors (MAAs) that lie outside minimal trap spaces are possible. 
MAAs are rare and (despite recent efforts) poorly understood. In this contribution to the  BMB \& JMB Special Collection ``Problems, Progress and Perspectives in Mathematical and Computational Biology'', we summarize the current state of knowledge regarding MAAs and present several novel observations regarding their response to node deletion reductions and linear extensions of edges. We conduct large-scale computational studies on an ensemble of 14~000 models derived from published Boolean models of biological systems, and more than 100~million Random Boolean Networks. Our findings quantify the rarity of MAAs (in particular, we found no MAAs in the biological models), but highlight the role of network reduction in introducing MAAs into the dynamics. We also show that MAAs are fragile to linear extensions: in sparse networks, even a single linear node can disrupt virtually all MAAs. Motivated by this observation, we improve the upper bound on the number of delays needed to disrupt a motif-avoidant attractor.}

\keywords{Boolean networks, Boolean models, discrete dynamics, complex systems, biomolecular networks, trap spaces, stable motif}



\maketitle

\section{Introduction}
Boolean networks are discrete dynamical systems that have been widely and successfully applied to model complex processes, especially in systems biology~\cite{schwab2020ConceptsBooleanNetwork,rozum2024boolean, abou-jaoude2016logical}.  They describe the causal logic of regulatory dynamics using \emph{nodes} to represent biomolecular entities (such as genes, proteins, or metabolites) that are connected by \emph{directed edges} representing regulatory influences. Each node can take on one of two states at any given time step---ON (1) or OFF (0)---that is updated according to an \emph{update function} of the node's incoming regulators' states. 

The nodes and edges that underpin the dynamics of a Boolean network form its \emph{interaction graph}. Signs are often assigned to the individual edges, indicating an activating (+1), inhibiting (-1), or ambiguous (0) influence. \footnote{An alternative representation of an ambiguous edge is a superposition of an activating edge and an inhibiting edge.} The interaction graph is the primary and best-constrained parameter of the Boolean network: numerous properties can be inferred solely from its structure~\cite{pauleve2012static}. However, many Boolean systems can share the same interaction graph, thus many problems cannot be conclusively solved based on the interaction graph alone. 

The dynamics of a Boolean network are affected by how nodes are selected for update in each time step~\cite{park2023models}. Various choices are possible, including updating all nodes (\emph{synchronous update}), randomly selecting a single node to update (\emph{asynchronous update}), or other more complicated schemes involving hidden states, such as the \emph{most-permissive} scheme of~\cite{pauleve2020reconciling}. In this work, we focus primarily (but not exclusively) on the asynchronous update scheme.

Typically, the long-term dynamics of a Boolean network (i.e., its \emph{attractors}) are intended to correspond to the biological phenotypes of the system under study. In the context of Boolean networks, attractors are defined as minimal trap sets~\cite{schwab2020ConceptsBooleanNetwork,rozum2024boolean} (i.e., the smallest sets of states that are invariant under the network's dynamics). There are various efficient methods to identify attractors, despite the NP-hardness of the problem~\cite{mori2022attractor,rozum2021parity,benes2021computing,trinh2021improved,balm}.

Attractors alone may be insufficient to answer many practical questions about Boolean networks. For example, one may wish to understand which interventions can or cannot drive the system toward a target behavior~\cite{paul2019efficient,cifuentes2022control,brim2023temporary}, which components of the system contribute to certain phenotypes~\cite{benes2022exploring}, or which initial conditions lead to which outcomes~\cite{albert2008boolean}. One fruitful approach toward answering these questions is to study how node states reinforce each other to lock a subnetwork of the system into a particular configuration (called a \emph{stable motif}~\cite{zanudo2013effective}). These can be identified, for example, using a hypergraph representation of the update functions~\cite{albert2003topology, zanudo2013effective, rozum2021parity}. Once reached, these stable configurations confine the dynamics to a subspace in which the constrained nodes are fixed, i.e., to a \emph{trap space}~\cite{klarner2015computing}. Stable motifs determine the system's commitment to one phenotype versus another. When a system has multiple stable motifs, the trap spaces nest within one another.

The most deeply nested \emph{minimal trap spaces} must each contain one or more attractors. However, there can be additional attractors that do not lie within any minimal trap space. In this case, the set of minimal trap spaces is called \emph{incomplete}~\cite{klarner2015approximating}, and an attractor outside all minimal trap spaces is called \emph{motif-avoidant}~\cite{rozum2021parity}, as states belonging to the attractor avoid the self-reinforcing network configurations (stable motifs) that result in constraining the dynamics to a minimal trap space. Despite recent efforts~\cite{richard2023attractor,naldi2023linear}, motif-avoidant attractors (MAAs) are still poorly understood, especially under asynchronous update. Empirical evidence suggests that motif-avoidant attractors are exceedingly rare in real-world biological networks~\cite{tonello2023attractor,balm}, but to the best of our knowledge, this observation has not been quantified before. 

In this work, we seek to lay out the current state of knowledge about MAAs and discuss some key open problems concerning their presence or absence. Inspired by~\cite{tonello2023attractor} and~\cite{naldi2023linear}, we explore how motif avoidance relates to \emph{node deletion reduction} and to adding a delay to a regulation (\emph{linear extension}). To gain further insight into these phenomena, we conducted multiple large-scale computational studies involving approximately one million networks derived from published biological models and one hundred million Random Boolean Networks (RBNs)~\cite{kauffman1969metabolic}. We use these insights to guide the development of some analytical results regarding MAA fragility.

In Sections~\ref{sec:preliminaries} and~\ref{sec:theoretical-results}, we introduce the necessary formal concepts and recall published results about motif-avoidant attractors. 
Section~\ref{sec:maas-defy-expectations} presents examples of MAAs that challenge our expectations of when motif avoidance should or should not be possible. 
Section~\ref{sec:maas-are-rare} explores the rarity of motif avoidance: We find no asynchronous MAAs in published Boolean models of biological systems, even when accounting for a wide array of model inputs and contexts. In RBNs, we show that the likelihood of encountering motif avoidance is non-trivial for dense networks (near $N=K$), but decreases rapidly for sparse networks. 
In Section~\ref{sec:maa-in-reductions}, we employ node deletion reduction and find several MAAs in the reducts of two published Boolean models. To the best of our knowledge, these are the first known MAAs in biologically interpretable asynchronous Boolean networks. We also demonstrate that the likelihood of encountering an MAA in a maximally reduced RBN does not depend on the size of the original (non-reduced) network. 
Finally, Section~\ref{sec:maa-with-delays} demonstrates that most MAAs can be eliminated by adding a delay node to a single edge, despite the known~\cite{naldi2023linear} requirement for $\mathcal{O}(|E|)$ delays in the worst case. We derive an improved upper bound for the number of delays needed to eliminate an MAA, present a family of networks that demonstrate that this bound cannot be lowered below $|E|/4$, and identify a simple criterion for when a single delay can destroy an MAA.

\section{Preliminaries}
\label{sec:preliminaries}

Boolean networks were introduced by Kauffman~\cite{kauffman1969metabolic} and Thomas~\cite{thomas1973boolean} as prototypical models for gene regulatory networks that underlie cell fate decisions (such as those that happen during cell differentiation). In the six decades since their introduction, Boolean models have proven effective in a variety of biological case studies, and the study of ensembles of generic Boolean networks (Random Boolean Networks) \cite{kauffman1969metabolic} yielded various hypotheses about cell dynamics and evolution.

\paragraph{Boolean function notation}
Any Boolean function can be expressed algebraically using the logical operators ``not'' (which we denote by ``\texttt{!}''), ``or'' (which we denote ``\texttt{|}''), and ``and'' (which we denote ``\texttt{\&}''). We denote variables with capital letters (e.g. \texttt{A}, \texttt{B}, \texttt{C}, \dots) and separate the variable and its function with ``\texttt{,}''. We use this notation due to its compactness and because it is frequently used in Boolean network analysis software. Example~\ref{ex:2d_MAA} demonstrates this notation on a simple two-variable Boolean model.

\paragraph{State transition graph}
Each Boolean system with $N$ variables induces a state transition graph (STG) with $2^N$ nodes that represent all possible system states. A directed edge from a node (system state) $S_1$ to another state $S_2$ exists when $S_1$ can be updated in one time step to obtain~$S_2$. Under synchronous update, the successor $S_2$ is obtained by applying \emph{all} update functions to $x$ in one step, hence each state has a single outgoing transition. Under asynchronous update, each state has between $0$ and $N$ outgoing edges, with each edge corresponding to the update of exactly one network variable (selected at random). Unless specifically noted otherwise, we assume the asynchronous update scheme throughout this work.

\paragraph{Trap spaces and succession diagrams}

A Boolean network \emph{subspace} is a set of states that is characterized by a subset of network variables fixed to constant values. In other words, a subspace is a hypercube in the state space of a Boolean network. Subspaces are typically denoted by $N$ symbols ``0'' (fixed to OFF), ``1'' (fixed to ON), or ``*'' (unconstrained), each giving the requirement for one variable. For example, the subspace denoted $0*1*$ corresponds to the set of states $\{ 0010,0011,0110,0111 \}$. A subspace is called a \emph{trap space} if its states form a trap set: a set that cannot be escaped by the network dynamics. Inclusion \emph{minimal trap spaces} are of particular interest here, as they represent parts of the state space that cannot be escaped, but that also cannot be further constrained into a smaller trap space. Importantly, in contrast to the state transition graph, trap spaces do not change with the chosen update scheme.

\begin{figure}
    \centering
    \includegraphics[width=.75\linewidth]{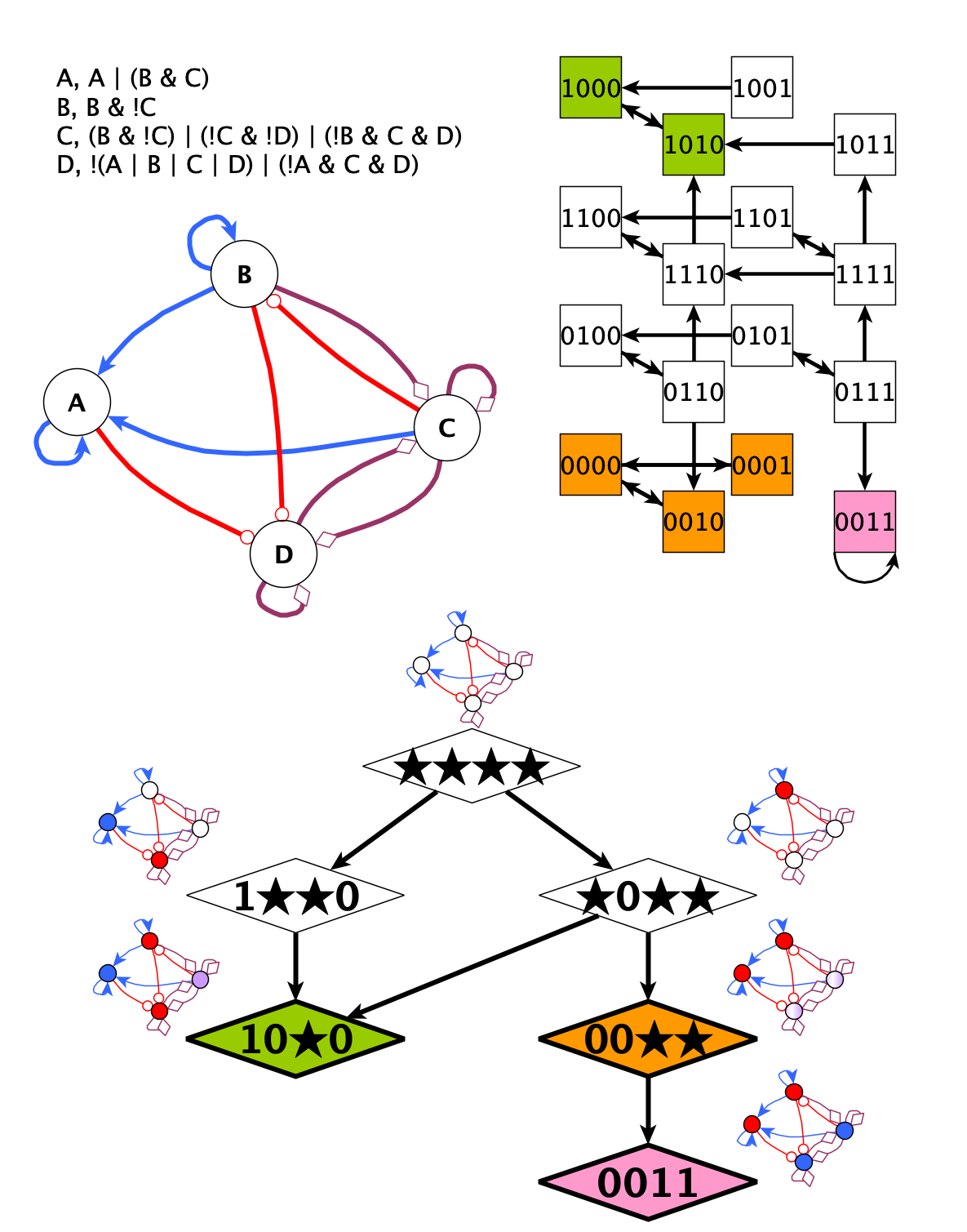}
    \caption{An example Boolean network that has a motif-avoidant attractor. The Boolean network's interaction graph and update functions are shown in the top left, with blue triangle-tipped arrows representing activation, red circle-tipped arrows representing inhibition, and purple diamond-tipped edges representing dual influences (either activation or inhibition, depending on context). The state transition graph of the system is shown in the top right, with attractors highlighted by color. On the bottom, the succession diagram is shown. Here, each node in this acyclic graph represents a trap space in the state transition graph, and is depicted alongside a cartoon of the network stable motif state it represents, with constrained nodes colored blue (for ON) or red (for OFF), and nodes that oscillate in purple. Edges denote set inclusion. The smallest trap space containing each attractor is colored accordingly; if that trap space is not a leaf node, i.e., a minimal trap space (as in the case of the orange attractor), the attractor is motif avoidant.}
    \label{fig:sd_mma_ex}
\end{figure}

Trap spaces can nest within one another, leading to a complex branching structure, called a \emph{succession diagram}~\cite{rozum2021parity} (see Figure~\ref{fig:sd_mma_ex}), that encodes which regulatory circuits (stable motifs) cause the system to commit to one fate or another. The succession diagram is a directed acyclic graph whose nodes are trap spaces (with fixed values percolated through the logical update functions) and whose edges represent inclusion. Thus, the leaf nodes of this graph are in one-to-one correspondence with the minimal trap spaces.

\paragraph{Motif-avoidant attractors}
The \emph{attractors} of a Boolean system are the minimal trap sets of the dynamics. In other words, they are the terminal strongly connected components of the STG. They are divided into two types: \emph{point attractors} (also called fixed points or steady states), which contain only one state, and \emph{complex attractors}, which contain more than one state (in synchronous update, these are always cycles; in asynchronous update, these can also be more complex components). A \emph{motif-avoidant attractor} (MAA) is a complex attractor (i.e., a minimal trap set) that does not lie within a minimal trap space. By this definition, the smallest subspace containing a motif-avoidant attractor is not a minimal trap space. Thus, either there exists a state in the subspace that has a transition out of the subspace, or the subspace contains a minimal trap space that is disjoint from the MAA. An example of the second case is given in Figure~\ref{fig:sd_mma_ex}.
Detecting MAAs is computationally difficult, and requires identifying both the attractors~\cite{mori2022attractor} and all minimal trap spaces~\cite{moon2022computational}.

\paragraph{Update schemes and their effect on attractors}
The states of nodes in a Boolean network can be updated in each time step according to various schemes. The earliest, and among the most popular, is the synchronous update scheme, in which all nodes are updated at every time step~\cite{kauffman1969metabolic}. While this scheme has computational and analytical advantages, it can also give rise to oscillations that depend on perfect synchronization of variables~\cite{park2023models}. This can be undesirable when these variables model inherently stochastic processes, such as gene transcription. The most popular alternative, and our focus in this work, is the general (also called stochastic) asynchronous update scheme, in which a single, randomly chosen node is updated at each time step. In the even more general random set update scheme, a randomly selected subset of the nodes is updated at each time step. The most permissive update scheme possible is the aptly named Most Permissive Boolean Network (MPBN) framework \cite{pauleve2020reconciling}, which introduces intermediate ``increasing'' and ``decreasing'' hidden states to allow a wider variety of transient behavior than is possible with other update schemes. All possible synchronous and asynchronous state transitions (and more) are permitted within the MPBN scheme. 

While the trap spaces (including fixed points) of a Boolean network are independent of the selected update scheme, the number and characteristics of oscillatory attractors do depend on how nodes are updated. Such attractors observed under one type of update may disappear when using a different update scheme. In particular, a Boolean network may have MAAs under one update scheme but not another.

In prior work, we have analyzed the impacts of update scheme in Boolean models from the literature~\cite{park2023models} and in random models~\cite{rozum2021parity}. We and others have observed that motif-avoidant attractors are fairly common in synchronously updated Random Boolean Networks~\cite{klemm2005stable}, and are a significant contributor to the substantially higher number of attractors observed under synchronous update~\cite{kaufman2005scaling} than under asynchronous update~\cite{greil2005dynamics, rozum2021parity}. They have also been observed in several models of cell processes~\cite{park2023models}. For example, a model of the cell cycle of neuroblastoma cancer cells~\cite{dahlhaus2016boolean} and a model of the interactions of 9 cell cycle transcription factors~\cite{orlando2008global} each exhibit a biologically substantiated motif-avoidant attractor under synchronous update; this attractor disappears under asynchronous update. This fragility to asynchronicity is typical of synchronous motif-avoidant attractors: the simple feedback loop system given in Example~\ref{ex:synchronous_MAA} illustrates this. 

\begin{example} [A synchronous motif-avoidant attractor destroyed by asynchronous update]\label{ex:synchronous_MAA}

The synchronous attractor $01\rightarrow10\rightarrow01$ is motif-avoidant (it avoids the trap spaces $00$ and $11$). This attractor vanishes under asynchronous update because changing the value of either node in $01$ or $10$ leads to one of the two point attractors. 
\begin{bnexample}
\begin{BVerbatim}
A, B
B, A 
\end{BVerbatim}
\end{bnexample}
\end{example}

Generally, the more transitions that are possible in a given update scheme, the less likely MAAs are to occur: in the extreme case, the MPBN framework ensures that MAAs do not occur at all. With respect to MAA occurrence, the asynchronous update scheme represents a widely used but poorly understood middle ground between synchronous update and the MPBN framework.

\paragraph{Node deletion reduction}

A variable of a Boolean network that does not self-regulate can be deleted by substituting its update function into the functions of the nodes regulated by it. The reduced Boolean system obtained after this deletion has been proven \cite{naldi2011dynamically,velizcuba2011reduction,tonello2023attractor} to preserve certain features of the original system's attractor repertoire. In particular, all point attractors are preserved, and complex attractors of the original system map to one or more complex attractors of the reduced system. Naldi et al. \cite{naldi2011dynamically} proved that any transition in the reduced system corresponds to at least one transition in the original system, and that the transitions starting from representative states (in which the update of the to-be-reduced node does not change its state) are preserved in the reduced system. Node deletion can be repeated while self-regulation-free variables exist in the system; the preservation of the point attractors and possible emergence of new complex attractors is maintained regardless of the number of deleted variables. It is possible that a complex attractor that emerges due to node deletion is a motif-avoidant attractor. 

\paragraph{Linear extensions}

Linear extension of an influence edge $\node{X} \to \node{Y}$ means the edge is replaced with an intermediary node $\node{Z}$ (network variable) and two edges $\node{X} \to \node{Z} \to \node{Y}$. The update function of $\node{Z}$ is simply $\node{X}$ (i.e. $\node{Z}$ propagates the value of $\node{X}$), and $\node{X}$ is replaced by $\node{Z}$ in the update function of $\node{Y}$. If such node $\node{Z}$ already exists in a Boolean network, we call it a \emph{linear node} (or a linear component)~\cite{naldi2023linear}. The deletion of a linear node is a special case of network reduction, thus the previously identified properties of network reduction apply to it. In particular, deleting a linear node may lead to the emergence of a motif-avoidant attractor. Furthermore, the addition of a linear node can eliminate a motif-avoidant attractor, and under certain well-defined conditions (as we discuss in more detail in Section~\ref{sec:theoretical-results}), this guarantees the network has no MAAs. 

\paragraph{Biological Boolean network ensemble}

To empirically evaluate the properties of motif-avoidant attractors in real-world models, we consider the \emph{Biodivine Boolean Models} (BBM) dataset~\cite{bbm}, which is to the best of our knowledge the largest collection of biologically relevant Boolean networks. The dataset covers 230 models aggregated from various sources, including The Cell Collective~\cite{cell-collective}, GINsim~\cite{ginsim}, Biomodels~\cite{biomodels}, the COVID-19 Disease Map project~\cite{covid-disease-map}, other literature reviews~\cite{kadelka2024meta}, and manually curated papers. Most of these models describe within-cell processes and cells' interaction with their environment or with pathogens. For each model, we infer the interaction graph from the semantic properties of the model's update functions, as some networks are originally published with redundant interactions or incomplete sign annotations. Most of these models contain input nodes, which denote information from the environment or factors outside of the scope of the model. As the values of input nodes significantly influence the attractor repertoire of Boolean models~\cite{park2023models}, we consider all input valuations when possible, which yields over $2^{100}$ model variants. In settings where this is impossible, we consider up to 128 unique, but randomly chosen valuations of these constant inputs, which yields 14~136 distinct model variants.

\paragraph{Random Boolean network ensembles}

We also consider a collection of N-K Random Boolean Networks (RBNs)~\cite{kauffman1969metabolic}, where each of the $N$ nodes receives $K$ edges from randomly selected nodes. The (quenched) update functions are chosen such that each of the $2^K$ input combinations yields an output of 1 with probability $p$. Tuning $p$ can produce an order-to-chaos transition~\cite{aldana2003Boolean} at $2Kp(1-p)=1$ (in the thermodynamic limit, when $N\rightarrow\infty$). Ensembles where $p$ satisfies this equation are called \emph{critical}. In this work, we study several large critical ensembles of RBNs across a wide range of $N$ and $K$.
Specifically, we consider ensembles of small RBNs with $N \in [4, \ldots, 10]$ together with the full range of admissible $K$, and large RBNs with $N \in [20,30,40]$ and $K \in [2, \ldots, 5]$. 

To achieve reasonable statistical significance for each ensemble (i.e. combination of $N$ and $K$), we sample random networks using \texttt{pystablemotifs}~\cite{pystablemotifs} until we discover 1000 networks with motif-avoidant attractors.\footnote{Sampling of the [N=40, K=4] and [N=40, K=5] ensembles had to be terminated prematurely after several weeks of runtime. The fact that these ensembles contain less than 1000 motif-avoidant cases is accounted for in the subsequent analysis.} Here, we are not concerned with the number of such attractors in individual networks, only whether \emph{at least one} motif-avoidant attractor exists. Overall, we tested more than 100 million random networks. The distribution of these networks across the ensembles is visualized in Figure~\ref{fig:random-ensemble-distribution} in the Appendix.

\paragraph{Tools and algorithms}

We provide a reproducibility artefact at \url{https://doi.org/10.5281/zenodo.13860057} which contains all code and data that is necessary to replicate the results of this paper. As stated previously, the artefact uses \texttt{pystablemotifs}~\cite{pystablemotifs} to generate the ensembles of critical RBNs. For structural manipulation of networks (reductions, linear extensions, etc.), we rely on \texttt{AEON.py}~\cite{aeon,aeon-py}. Finally, to detect motif-avoidant attractors and their associated trap spaces, we use the symbolic techniques implemented in \texttt{AEON.py} in combination with the succession diagrams generated by \texttt{biobalm}~\cite{balm}. \texttt{AEON.py} uses binary decision diagrams (BDDs) to symbolically encode the STG of a Boolean network, such that even large state spaces (more than 30 variables) can be explored efficiently. Furthermore, BDDs can be used to analyze and transform the BN update functions, for example, to obtain the underlying interaction graph or to simplify a BN after a reduction. Meanwhile, \texttt{biobalm} uses answer set programming to generate the network's succession diagram, which can be used to identify motif avoidance, enclosing trap spaces of motif-avoidant attractors, but also to speed up attractor computation in general.

\section{Theoretical results relevant to motif-avoidant attractors}
\label{sec:theoretical-results}

As proven by A. Richard \cite{richard2010negative}, under asynchronous update, the existence of a non-positive cycle (determined by the product of edge signs) in the interaction graph is a requirement for a complex attractor (sustained oscillation). In addition, in order to have a minimal trap space that is not trivially the entire state space, a Boolean network must have at least one non-negative cycle in its interaction graph. By similar logic, a network needs to have multiple attractors in order to have a motif-avoidant attractor, which also requires the presence of a non-negative cycle \cite{richard2007necessary}. Thus, Boolean networks with no cycles, only positive cycles, or only negative cycles cannot have motif-avoidant attractors (MAAs).

\subsection{Separating and trap-separating networks}

Richard and Tonello \cite{richard2023attractor} derive necessary conditions for the interaction graph of a Boolean network such that the network is not \emph{separating} (i.e., there is a pair of attractors whose smallest subspaces overlap) or not \emph{trap-separating} (i.e., there is a pair of attractors whose smallest trap spaces overlap). A motif-avoidant attractor and the attractor determined by the avoided motif are not trap-separating, and may not be separating (i.e., they could be in the same subspace, as we will see in examples in later sections). Richard and Tonello prove that if the Boolean network is not trap-separating, then its interaction graph has a path from a non-positive cycle to a non-negative cycle. They also prove that if the Boolean network is non-separating, three statements are true about the interaction graph: it has either a non-negative cycle that intersects a non-positive cycle or a cycle in which at least one edge is ambiguous, it has at least two non-positive cycles or an ambiguous cycle with at least two ambiguous edges, and at least two vertices must be removed to destroy all the non-negative cycles.   

\subsection{Boolean networks with L-cuttable interaction graphs}

The interaction graph of a Boolean network admits a linear cut when a linear node (a node with in-degree and out-degree one) occurs in each cycle (including self-edges) and in each path from a node with multiple targets to a node with multiple regulators. Naldi et al. \cite{naldi2023linear} proved that in a Boolean network whose interaction graph admits a linear cut the attractors are in a one-to-one relationship with minimal trap spaces. As a consequence, these L-cuttable networks do not have motif-avoidant attractors. In addition, in an L-cuttable network, all states reachable from a canonical state (a system state in which each linear node has the same state as its sole regulator) by updating any set of nodes are also reachable by updating one node at a time. Naldi et al. propose that the unreachability of certain states under asynchronous update is due to unsatisfiable circular requirements on the order of update of the nodes. An example of such an unsatisfiable circular requirement is ``In order to reach system state $Y$ from system state $X$, node $\node{A}$ needs to be updated before $\node{B}$ (because the updated state of $\node{A}$ is needed to allow the state change of $\node{B}$), $\node{B}$ needs to be updated before $\node{C}$, and $\node{C}$ needs to be updated before $\node{A}$ (because the un-updated state of $\node{A}$ is needed to allow the state change of $\node{C}$)''. Naldi et al. observe that such unsatisfiable constraints arise from feedback loops or incoherent feed-forward loops in the interaction graph. These conflicts can be resolved by adding linear extensions to edges until the interaction graph admits a linear cut. In the worst case, all the edges need linear extensions. 

\subsection{Linear extensions}

Each linear node added in a linear extension of an edge can be thought of as a delay node, an intermediary between its parent node (the starting node of the original edge) and the end node of the original edge. Each delay node serves as a memory of the un-updated state of its parent node. If each edge has a delay node, both states of each node are represented in the system, thus unsatisfiable requirements on the update order are no longer possible. 

A key feature of a motif-avoidant attractor is the unreachability of certain state(s) of the enclosing subspace from the states visited by the attractor. While a maximal extension is sufficient to eliminate any motif-avoidant attractor by ensuring the reachability of all states in the subspace, the gain in reachability needed to eliminate a specific motif-avoidant attractor may be achieved with fewer extensions. Consider a system with a 2-variable MAA combined with a minimal trap space $11$.

\begin{example} [Prototypical 2-variable motif-avoidant attractor]\label{ex:2d_MAA}
The sole network with two variables that exhibits a motif-avoidant attractor.

\begin{bnexample}
\begin{BVerbatim}
A, (!A & !B) | (A & B)
B, (!A & !B) | (A & B)
\end{BVerbatim}
\end{bnexample}
\end{example}

In Example \ref{ex:2d_MAA} the reachability of the state $11$ from the state $00$ has an unsatisfiable requirement: node $\node{A}$ needs to be updated before node $\node{B}$ to allow its state change, and node $\node{B}$ needs to be updated before node $\node{A}$ to allow its own state change. A full linear extension of this system requires 4 delay nodes. Yet, the unsatisfiable requirement can be eliminated by a single delay node. For instance, when adding a delay node $\node{C}$ to the $\node{A} \to \node{B}$ edge, the requirement for the reachability of $111$ from $000$ becomes ``$\node{A}$ needs to be updated before $\node{B}$, and $\node{B}$ needs to be updated before $\node{C}$'', which is now satisfiable due to the newly added delay node. 

The literature discussed in this section has produced various necessary conditions for the elimination of motif avoidance; these conditions are often quite strict, and are not met for many networks that do not contain MAAs. Motivated by this observation, we sought to investigate the related questions of what makes motif avoidance appear and what is the minimal number of delays that eliminates a motif-avoidant attractor. Our investigation was small-scale at first. We performed an exhaustive search to enumerate the possible MAAs of 3-variable Boolean networks. We dedicated a significant effort to identifying conditions that guarantee or rule out motif avoidance. In the next section, we present a sampling of examples to illustrate the challenges of this investigation.

\section{Motif-avoidant attractors defy expectations}
\label{sec:maas-defy-expectations}

 We formulated various conjectures about the requirements for motif avoidance and then found counterexamples for each conjecture. The examples in this section are representative of natural expectations regarding motif-avoidant attractors that are then contradicted by reality. The interaction graphs and state transition graphs of these systems are shown in Figure \ref{fig:example_figures} in the Appendix.  To focus on the MAA in these examples, we combine them with a single-state minimal trap space (i.e., a point attractor). In general, the MAA can appear at any level of a succession diagram, in which case these examples would be parts of larger systems. 

\begin{example} [Simple functional forms]\label{ex:monotonic_nested} This example shows that MAAs can exist even in Boolean networks with locally monotonic update functions.

\begin{bnexample}
\begin{BVerbatim}
A, (!A & !B) | C
B, (!A & !B) | C
C,  (A &  B)    
\end{BVerbatim}
\end{bnexample}

This system has a minimal trap space $111$ and a motif-avoidant attractor that visits 3 states of the $**0$ subspace. The fourth state of the subspace, $110$,  has a transition to the minimal trap space $111$.
\end{example}

It is well-documented that a regulator’s direct effect on a target tends to be monotonic in biological networks, and that multiple regulators tend to form a hierarchy in which strong (canalyzing) regulators override the weaker regulators’ effect, forming nested canalyzing functions~\cite{harris2002model, jarrah2007nested, subbaroyan2022minimum, kadelka2024meta}. The example above shows that these strong restrictions on the update functions do not rule out motif-avoidant attractors. 

\begin{example} [No need for negative self-edges]\label{ex:template2iv} Although negative cycles are necessary to have MAAs, negative self-edges are not.

\begin{bnexample}
\begin{BVerbatim}
A, !C | (A & B)
B, !A | (B & C)
C, !B | (A & C)
\end{BVerbatim}
\end{bnexample}

This Boolean network has a minimal trap space $111$ and a motif-avoidant attractor that forms a cycle of 6 states. This MAA would be preserved under synchronous update as well as random set update, because the functions do not allow the state change of multiple nodes. 
\end{example}

Furthermore, it is not true that the minimal trap space is always reachable from the non-attractor states of the MAA subspace. That is, to escape the motif-avoidant attractor, it may be necessary to perturb the network to a state outside the MAA subspace:

\begin{example} [Temporary escape from the subspace]\label{ex:out_and_in_transition} One way in which a subspace that contains an MAA can be non-trapping is the existence of a trajectory that leaves and then reenters the subspace.
\begin{bnexample}
\begin{BVerbatim}
A, (!A & !B & !C) | (A & C) | (A & B) 
B, ( A & !B & !C) | (A & B & C) | (!A & B & !C)
C, (B & C) | (!A & B)
\end{BVerbatim}
\end{bnexample}

The Boolean network has a minimal trap space 111 and a motif-avoidant attractor that visits the states $000$, $100$, and $110$. The fourth state of the $**0$  subspace, $010$, is the starting point of the path $010 \to 011 \to 001 \to 000$ (see Figure \ref{fig:example_figures} in the Appendix).
\end{example}

Finally, just as multiple asynchronous attractors can appear in one minimal trap space, a single subspace can contain more than one motif-avoidant attractor. That is, the smallest trap space containing an MAA may also contain an additional MAA:

\begin{example} [Two independent MAAs can coexist in the same enclosing trap space]\label{ex:two_MAAs}

\begin{bnexample}
\begin{BVerbatim}
A, (!A &  B & !C) | (A & !B) |  (A & C)
B, (!A &  B & !C) | (A & !B) | (!B & C) | (A & C)
C, (!A & !B & !C) | (B &  C) |  (A & C)
\end{BVerbatim}
\end{bnexample}

The Boolean network has a minimal trap space $111$, a motif-avoidant attractor that visits the states $000$, $001$, $011$, and a second MAA that visits the states $100$, $110$, $010$.
\end{example}

Seeing the diversity of motif-avoidant attractors in small Boolean networks, we followed up with extensive computational studies of Boolean network ensembles, as described in the next section.

\section{Motif-avoidant attractors are surprisingly rare}
\label{sec:maas-are-rare}

Though motif avoidance in asynchronously updated Boolean networks is known to be more rare than in synchronously updated networks (see Section~\ref{sec:preliminaries}), the frequency of asynchronous motif avoidance has not previously been studied systematically. Here, we quantify how common MAAs are under asynchronous update in two widely-used network ensembles.

\subsection{Motif avoidance in real-world Boolean networks} 

We performed an extensive computational study of published Boolean models, analyzing all models from the BBM dataset~\cite{bbm}. Note that most of these models contain input nodes with constant values. In many cases, only a subset of the exponentially many input valuations was explored by the model's authors, and many models have not been checked for motif-avoidant attractors at all. Even though the model behavior has not always been biologically confirmed across all input valuations, here, we still consider all input valuations as relevant, since they represent models that can be linked to real biological processes.

When possible, we use the symbolic ``colored'' representation in \texttt{AEON.py}~\cite{aeon,aeon-py} to check all models in the BBM dataset across all input valuations for MAAs. This technique was successful for 221/230 models, accounting for more than $2^{100}$ input valuations overall. For the remaining models, we use randomly sampled unique input valuations (see Section~\ref{sec:preliminaries}) and check them for MAAs using \texttt{biobalm}~\cite{balm}, which scales better for large networks when the input values are fixed. Together, this accounted for 1004/1026 input valuations across the remaining 9/230 models.\footnote{The remaining 22 models are all variants of a single BBM model that can, depending on the input values, exhibit a very large number of minimal trap spaces that could not be fully enumerated even with a one week timeout.} \emph{No motif-avoidant attractor was found in any of these models.}

We tested whether the lack of MAAs is due to L-cuttability. Only 9/230 models in the BBM dataset have an L-cuttable interaction graph when considering input nodes as variables, but ignoring their self-regulations, as these alone cannot cause MAAs to appear. To evaluate the simplification due to fixed input values, we analyzed up to 128 random constant valuations for each model and tested their L-cuttability after the constant values were percolated into their target functions. The interaction graphs are indeed greatly simplified: 5386/14136 of the instances percolate to an empty graph, and 1277 of the remaining 8750 non-trivial graphs are L-cuttable. Taking the effect of constant nodes into account thus greatly increases the number of L-cuttable interaction graphs (to 47\%), but not enough to account for the complete lack of MAAs.

\subsection{Motif avoidance in random N-K networks}
\label{sec:motif-avoidance-in-random-networks}

To expand the search, we also considered ensembles of critical Random Boolean Networks according to the N-K model~\cite{kauffman1969metabolic}, sampled using \texttt{pystablemotifs}~\cite{pystablemotifs}. These ensembles are described in more detail in Section~\ref{sec:preliminaries}.

\begin{figure}
    \centering    
    \includegraphics[width=1.0\linewidth]{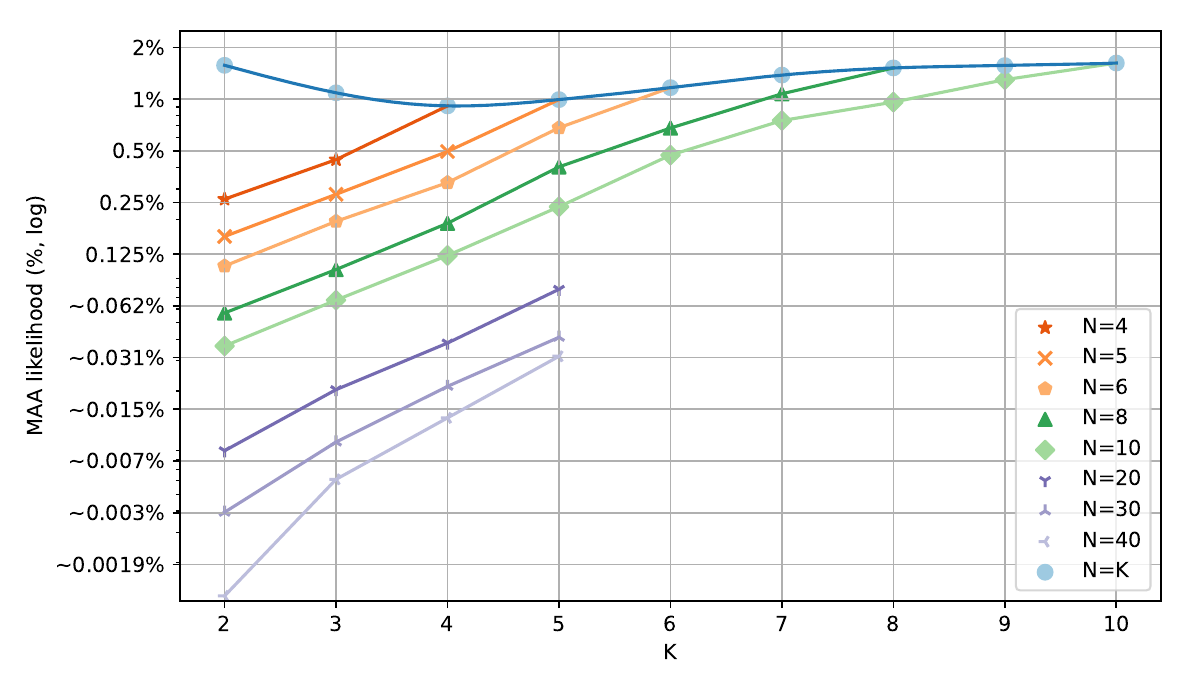}
    \caption{The likelihood of encountering a network with a motif-avoidant attractor within the ensembles of critical N-K random Boolean networks. Note that the likelihood axis is logarithmic. The blue trend-line shows the likelihood for maximally dense networks where $N=K$, the remaining trend-lines show fixed network sizes with varying $K$ (for large networks, the data only cover $K \in [2,3,4,5]$). The 95\% confidence intervals for all points are smaller than the displayed points.}
    \label{fig:random-networks-maa-likelihood}
\end{figure}

Figure~\ref{fig:random-networks-maa-likelihood} shows the likelihood of encountering a network with a motif-avoidant attractor in each of the tested ensembles (excluding $N=7$ and $N=9$ for better visual clarity). Notice that even for dense networks of non-trivial size (e.g. $N=K=10$), the MAA likelihood is only $\sim1.6\%$. Furthermore, the maximum MAA likelihood for each $N$ is achieved when $N=K$. For a fixed value of $K$, the MAA likelihood decreases very quickly with growing~$N$. For $K=2$, the MAA likelihood is less than $0.04\%$ when $N \geq 10$ and well below $0.001\%$ for $N \geq 20$. Biological networks are known to be sparse and contain at least tens of nodes (in the BBM dataset, the average $K$ is $\sim 2.3$ and the average $N$ is $\sim 58.01$). Therefore, a contributor to the lack of MAAs in the BBM models may be that in sparse networks with asynchronous update, the possibility of creating an MAA by chance is simply too small. With 230 samples and no observations of motif avoidance, the 95\% confidence interval for the proportion of biologically realistic models with MAAs is $[0, 1.3\%]$.

\section{Intensifying the search for motif-avoidant attractors by reducing networks}
\label{sec:maa-in-reductions}

As can be seen in our results presented in Section \ref{sec:motif-avoidance-in-random-networks} and the examples above, all the motif-avoidant attractors we have identified involve a dense network or subnetwork. Such dense subnetworks are very rare in Boolean models. As we did not find any MAAs in the published Boolean models, we propose that reducing networks (by deleting nodes, as described in the Preliminaries) may be one way that MAAs may arise in practice. This proposition is supported by the previous result \cite{naldi2023linear} that the deletion of linear nodes destroys L-cuttability, thus enabling motif avoidance. Yet, we found that node reduction can also eliminate a motif-avoidant attractor by transforming it into a regular complex attractor. We start this section with examples, then we present the results of our search for MAAs in reduced versions of published Boolean models and in maximally reduced Random Boolean Networks.

\subsection{Motif avoidance is a non-local property: node reduction can either create or eliminate it}

Node deletion reduction preserves point attractors, and complex attractors of the original system map to one or more complex attractors of the reduced system \cite{naldi2011dynamically, velizcuba2011reduction}. It is possible that a complex attractor created by node deletion is a motif-avoidant attractor.

\begin{example} [Reducing a linear node can create an MAA]\label{ex:template1_delay} The MAA created by reducing node $\node{C}$ is identical to that of Example \ref{ex:2d_MAA}. It visits three states and avoids the minimal trap space $11$.
\begin{bnexample}
\begin{BVerbatim}
A, (A & C) | (!A & !C)
B, (A & B) | (!A & !B)
C, B
\end{BVerbatim}
\end{bnexample}
\end{example}

Conversely, we found that node deletion reduction can lead to the emergence of a stable motif, which traps a motif-avoidant attractor into a minimal trap space. 

\begin{example} [Reduction can eliminate motif avoidance] \label{ex:reduction} The Boolean network below has a minimal trap space $111$ and a motif-avoidant attractor that visits 6 states. 
\begin{bnexample}
\begin{BVerbatim}
A, C
B,  (A & !B & !C) | (A & B & C)
C, (!A & !B & !C) | (A & B & C)
\end{BVerbatim}
\end{bnexample}
When reducing node $\node{A}$, the term {\upshape\texttt{(C \& !B \& !C)}} in the function of $\node{B}$ evaluates to 0.
\begin{bnexample}
\begin{BVerbatim}
B,  B & C
C, (B & C) | (!B & !C)
\end{BVerbatim}
\end{bnexample}

The reduced Boolean network has a new stable motif $\node{B}=0$, which defines a minimal trap space $0*$. There no longer is a motif-avoidant attractor. 
\end{example}

The reduced Boolean network's interaction graph is complete. Only by linear extension of all 4 edges would the interaction graph become L-cuttable. Thus, Example \ref{ex:reduction} illustrates that L-cuttability is not a necessary condition for the absence of a motif-avoidant attractor. Furthermore, the interaction graph of the reduced system satisfies the three criteria identified by Richard and Tonello~\cite{richard2023attractor} as necessary for a Boolean network to be non-separating. Yet, the system is in fact separating: there is a point attractor $11$ and a complex attractor that fills the $0*$ subspace. The lack of motif avoidance in the reduced system suggests that the currently known criteria for identifying it are too strict.

The reduced system of Example \ref{ex:reduction} also illustrates that an attractor that has a single oscillating node and has fixed states for all other nodes corresponds to a minimal trap space. Two nodes must oscillate for a motif-avoidant attractor to exist.

In the examples we encountered, we observed that the MAA-trapping stable motif emerges due to the elimination of a node's dependence on a regulator. This happens because newly coincident influences yield a constant due to a logical identity (such as \texttt{X | !X $\equiv$ 1}).

\subsection{Reduction introduces motif-avoidant attractors to biological networks}
\label{sec:reduction-in-bbm-networks}

Knowing that node deletions eliminate L-cuttability and generally increase network density, our goal is now to test whether node deletions increase the MAA likelihood in biological networks. However, as we showed with the previous examples, node deletions can both create and destroy MAAs. To reflect these possibilities, we set up our numerical experiment as follows:

For each model from the BBM dataset, we test all input valuations if there are fewer than seven inputs or sample up to 128 unique constant input valuations. This process results in 14~136 model variants. Then, for each of these model variants, we generate reduced Boolean networks by eliminating variables one by one until no further elimination is possible (i.e. each remaining variable has a self-regulation). To select the variable for elimination, we prioritize those with simple update functions (i.e. small binary decision diagram representations). We then output each such network as a separate test instance, resulting in 1~157~940 Boolean networks. Out of these models, we were able to analyze $\sim$~750~000 within a timeout of 10 minutes per model. In particular, the results account for all models with fewer than 78 nodes. Based on Section~\ref{sec:motif-avoidance-in-random-networks}, we would expect MAAs to be increasingly rare in larger models, which is why we have not pursued a greater timeout.

Among these Boolean networks, we identified 9 MAAs with between 4 and 10 oscillating variables. When mapping these nine instances to the original, non-reduced models, they correspond to two models: BBM-151, originally published in \cite{sanchez2019contribution} and BBM-202, originally published in \cite{gupta2022boolean}. Later in Section~\ref{sec:maas_in_published_models}, we provide a detailed discussion of what causes these MAAs to appear in these models and explore their biological relevance.

We should also note that while we made a substantial effort to consider a wide range of model configurations and their reductions, we did use a deterministic method to select which variable is reduced. The order in which variables are reduced does matter in many cases. For example, after identifying the root cause of motif avoidance in models BBM-151 and BBM-202, we could manually derive other reduced variants of these models (often by applying fewer reductions) that admit MAAs. As such, our results cannot be considered exhaustive. Nevertheless, the total number of possible reductions across all BBM models is astronomical, so with this numerical experiment, we hope to provide an illustration of how rare motif-avoidant attractors are even in reduced networks. The emergence of these motif-avoidant attractors motivates further study into the relationship between node deletion and motif avoidance.

\subsection{Reduction introduces motif-avoidant attractors to sparse random networks}

\begin{figure}
    \centering    
    \includegraphics[width=1.0\linewidth]{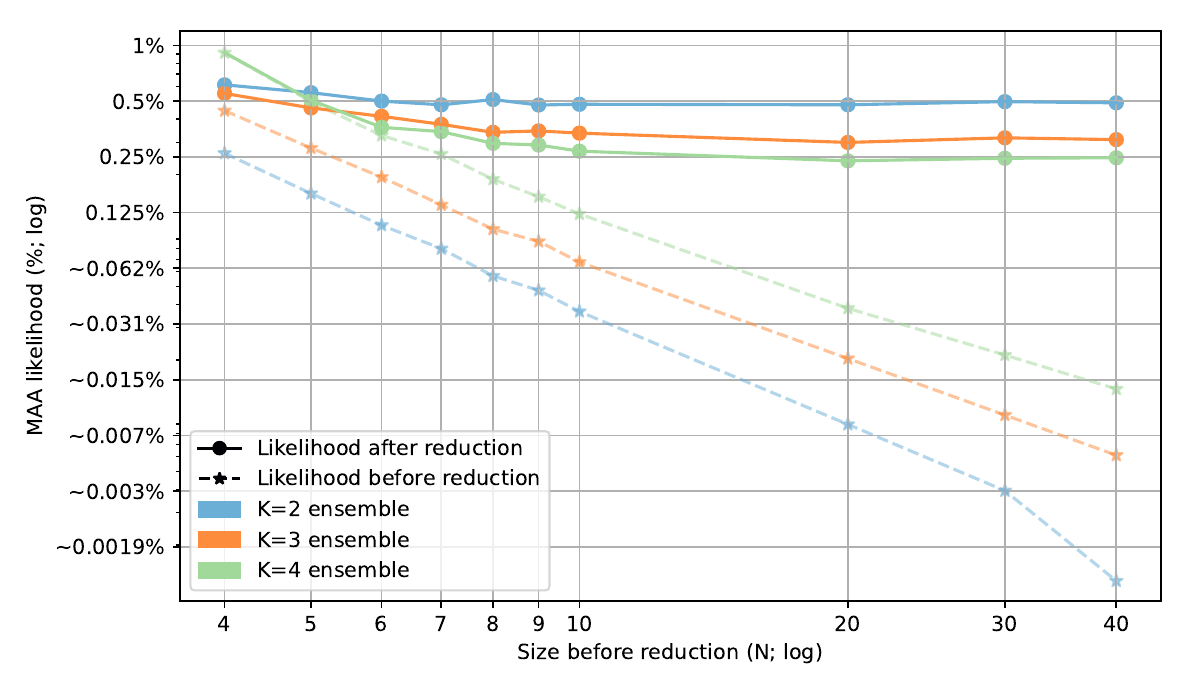}
    \caption{The likelihood of encountering a motif-avoidant attractor \emph{after} maximally reducing the networks (solid trend-line) from the random critical N-K ensembles (see Figure~\ref{fig:random-ensemble-distribution}) compared to the likelihood of the full networks before reduction (dashed trend-lines). Note that both axes are logarithmic.}
    \label{fig:random-networks-reduced-maa-likelihood}
\end{figure}

As we have seen, node deletion can, in practice, lead to the introduction of MAAs. To quantify this effect, we again turn to the critical Random Boolean Networks. We take the random networks sampled in Section~\ref{sec:motif-avoidance-in-random-networks} and we reduce each network according to the process outlined in Section~\ref{sec:reduction-in-bbm-networks}. However, in this case, we are only interested in the final, maximally reduced Boolean network, where no further reductions are possible. We then test each such maximally reduced network for MAAs and record the results, just as in Section~\ref{sec:motif-avoidance-in-random-networks}. We stop once 1000 fully reduced Boolean networks with motif avoidance are observed in each ensemble.

Based on this data, we can compare the likelihood of encountering an MAA in a sampled random Boolean network to the MAA likelihood if said network were to be fully reduced. This trend is visualized in Figure~\ref{fig:random-networks-reduced-maa-likelihood} (the case of $K=5$ was measured, but is omitted for visual clarity). We see that in a set of original (non-reduced) networks with a fixed value of $K$, the MAA likelihood decreases approximately exponentially with $N$. However, if these networks are fully reduced, their MAA likelihood converges to a constant value, and does not decrease meaningfully even for relatively large networks (i.e., for N=40).

This result could be partially attributed to the fact that in sparse networks, the node deletion reduction is extremely effective at decreasing the overall network size: In the case of $K=2$, the average size of the reduced network is $2.0$ for an original network with N=10 and $2.7$ for N=40. However, if these reduced networks were random, even such a small difference in size ($2.0$ versus $2.7$)  should result in a meaningfully different MAA likelihood (see Figure \ref{fig:random-networks-maa-likelihood}). Yet, the observed likelihood is $0.48\%$ in both cases. For larger $K$, the differences in size are even greater: For $K=5$, we have an average fully reduced size of $5.9$ for N=20 and $8.9$ for N=40, yet the likelihood is the same at $0.22\%$. 

We observe that the average number of interactions per variable in these reduced networks also increases with $N$ (and thus with the degree of network reduction): In our first example, it goes from $1.25$ to $1.43$, while in the second example, it increases from $4.5$ to $6.9$ interactions per variable. This suggests that the MAA likelihood for the fully reduced networks is kept constant by the interplay between network size and the number of interactions. Even though larger networks cannot be reduced as much as the smaller ones, the number of interactions per variable in the larger reducts is (perhaps counter-intuitively) also greater. Together, our results suggest that these two effects cancel each other out exactly (larger networks have a smaller MAA likelihood, but more interactions increase the MAA likelihood), resulting in the MAA likelihood staying constant.

Our analysis demonstrates that even though motif-avoidant attractors are extremely rare in both biological and random networks, network reduction does meaningfully alter the likelihood of their presence. Furthermore, at least for critical random networks with a fixed $K$, the MAA likelihood of a maximally reduced network does not actually decrease with the original $N$. Modelers always face the question of what resolution to use when constructing a model of a biological process. Should every gene product be included in the model, or can linear pathways be shortened into edges? Our results caution against the indiscriminate use of compression during modeling.

\section{Investigating the robustness of motif avoidance to delays}
\label{sec:maa-with-delays}

After having documented the increase of the MAA likelihood caused by network reduction, we turn to the opposite process: network expansion by linear extension of edges. Naldi et al. \cite{naldi2023linear} identified a contributing factor to motif avoidance in the unreachability of certain states under asynchronous update due to unsatisfiable requirements on the order of update of the nodes. Although L-cuttability (achieved by linear extension of up to all edges in the network) guarantees the lack of MAAs, we found that MAAs are rare even in non-L-cuttable networks. This finding leads us to propose that motif avoidance can be eliminated by fewer linear extensions than those required by L-cuttability. We evaluate this proposition by investigating two limiting cases. We identify (theoretically and numerically on the MAAs in our collection) cases when a motif-avoidant attractor is eliminated by linear extension of a single edge. We also present leads toward lowering the upper bound for the number of delays needed to eliminate an MAA.

The basis of the elimination of a motif-avoidant attractor is the property that any state that is outside a trap space but can reach the trap space cannot be part of an attractor. Thus, if linear extension of a single edge (which adds a delay variable and doubles the number of states in the Boolean network) makes the states derived from the original system's MAA reach the trap space, the delay-expanded system no longer has an MAA. Linear extension of an edge is the reverse of the deletion reduction of a linear node. Thus, we can apply the relationships between the trajectories of a reduced Boolean network obtained by deleting a self-edge-free node and the trajectories of the original network \cite{naldi2011dynamically} to derive reachability relationships in a system in which a single edge is extended by a linear node. In this context, we call the states referred to as `representative states' by Naldi et al. `canonical states'.
 
Paraphrasing Lemma 1 of \cite{naldi2011dynamically} yields information on transitions that start from canonical states. First, for every state transition $S_1\rightarrow S_2$ in the original system, there will be a transition starting from the canonical state corresponding to $S_1$ in the delay-expanded system. Second, for every transition in the delay-expanded system that starts from a canonical state, there is a corresponding transition in the original system (i.e., when not considering the state of the delay node). This means that the canonical states retain the connectivity of the original system. 

A node affected by a delay can have its update function evaluated to more than one value for a given configuration of the original system, depending on the value of the delay node. As a consequence, non-canonical states (memory states) can have transitions that the original system did not have, or can lack transitions that the original system had. We found that lost transitions can interfere with the ability of gained transitions to eliminate the motif-avoidant attractor (see Figure \ref{fig:not_all_delays_work} for example). Gained transitions of memory states are well-characterized by application of previous results by Naldi et al. \cite{naldi2011dynamically}, as we will describe next. In contrast, lost transitions of memory states have not been previously studied (to our knowledge). We present a brief overview of lost transitions in Appendix \ref{sec:supplementary_text}.

It has been previously shown (Property 1 of \cite{naldi2011dynamically}) that a transition of the delay-expanded system is missing from the original system if and only if (i) the transition starts from a memory state, and (ii) the transition updates and changes the value of the delay-affected node. Each state of the delay-expanded system that reaches the trap space due to one of these new transitions cannot be part of an attractor. If the states derived from the states of the original system's MAA reach the trap space due to one of these new transitions, the delay-expanded system no longer has an MAA. 

The results concerning the connectivity of canonical states can be generalized from a single to multiple linear extensions. The canonical states of a delay-expanded system can have more connectivity compared to the original system (due to the transitions of memory states that are different from the transitions of the original system), but never less. Furthermore, a Boolean system with multiple delays always retains any connectivity of the canonical states that exist in the Boolean systems that have a subset of those delays. 
We use these properties to visualize the new transitions enabled by linear extensions through an economical projection of state transition graphs. 

\subsection{The projected state transition graph}

\begin{figure}
    \centering
    \includegraphics[width=1\linewidth]{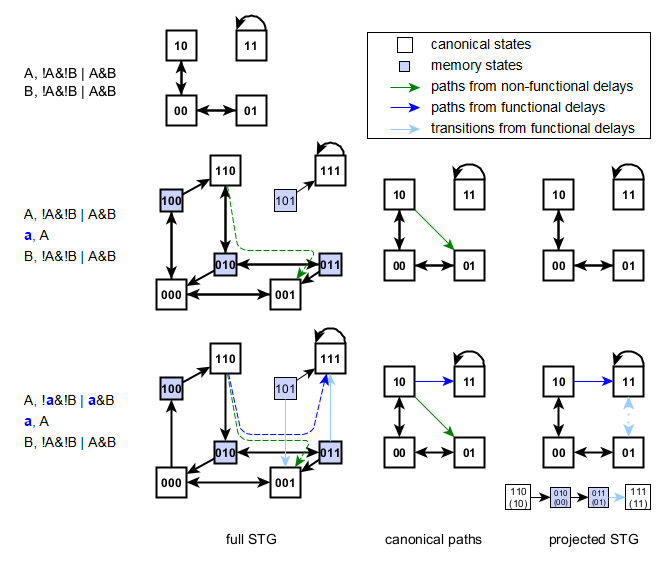}
    \caption{Illustration of linear extensions and projected state transition graphs (STGs) in case of Example \ref{ex:2d_MAA}. Each panel indicates the functions and STG of the original system (top), a system with a non-functional delay node (middle), and a system with linear extension on the self-edge of $\node{A}$ (bottom). The states and state transitions that make up the attractor of the original system are shown in bold. Memory states, in which the state of the delay node $\node{a}$ is different from the state of its parent node $\node{A}$, are marked by a blue background. The light blue edges of the STG show transitions that occur due to the functional delay node. A path in the full STG starting from a canonical state, passing through one or more memory states, and ending in another canonical state is shown in dashed lines, and is projected to an edge among the original system's states in the second column. Paths that appear in the system with the non-functional delay node are shown with green, and paths that appear only in the system with functional delay are shown with blue. The projected STG adds to the STG of the original system the projections of the new transitions allowed by the functional delay node (i.e., the blue edges). Each succession of state transitions that corresponds to a blue edge of the projected STG is indicated below the projected STG; it indicates the corresponding state of the delay-free system in parenthesis and highlights memory states with a light blue background. The projections of the transitions due to the functional delay node (i.e. the projections of the light blue edges) are shown with light blue dotted lines in the projected STG.}
    \label{fig:projected_stg_intro}
\end{figure}

To effectively illustrate the mechanism by which a linear extension eliminates a motif-avoidant attractor, we propose a mapping of the delay-extended system's state transition graph to the original system's state transition graph (STG). We call this mapping the projected state transition graph. In the projected state transition graph a pair of states of the original system is connected by a directed edge if the two corresponding canonical states of the delayed system are connected by a path that does not involve any other canonical states. Note that if a sink node with a single regulator is added to the system, it will create paths among canonical states that do not contain any other canonical states. However, we do not include such paths as edges of the projected STG to focus on the functional effects of the delay node. This choice is similar to the choice of not including in a state transition graph the self-transitions that correspond to a node being updated but not changing state. Figure \ref{fig:projected_stg_intro} illustrates the method of deriving the projected state transition graph of Example \ref{ex:2d_MAA}. The bottom row of Figure \ref{fig:projected_stg_intro} also offers two examples of transitions starting from memory states that are not present in the original system nor for canonical states: $101 \rightarrow 001$ and $011 \rightarrow 111$. We indicate such transitions with blue lines in the delay-expanded system's STG and with light blue dotted lines in the projected STG.

\subsection{Motif-avoidant attractors in small, dense networks can be destroyed by single delays}

In our analysis of the motif-avoidant attractors of small, dense networks, we found that single delay nodes (i.e., linear extensions of single edges) were successful in eliminating the MAA. Indeed, a delay to any of the four edges of the 2-variable system of Example \ref{ex:2d_MAA} will destroy the MAA. In the system of Example \ref{ex:monotonic_nested}, linear extension to any of four edges, i.e. $\node{A}$ to $\node{B}$, $\node{A}$ to $\node{C}$, $\node{B}$ to $\node{A}$, $\node{B}$ to $\node{C}$, will eliminate the MAA. In the system of Example \ref{ex:template2iv}, linear extension to the self-regulation of any of the three nodes will destroy the MAA. The MAA of Example \ref{ex:out_and_in_transition} is destroyed by linear extension of the edge from $\node{A}$ to $\node{C}$. The motif-avoidant attractor of Example \ref{ex:reduction} can be eliminated by linear extension to any of four edges, namely the self-regulation of $\node{B}$, the self-regulation of $\node{C}$, $\node{B}$ to $\node{C}$ or $\node{C}$ to $\node{B}$.

For a more systematic investigation of the frequency of MAAs that can be eliminated by a single delay, we aimed to identify all types of MAAs in 3-variable systems. We considered a trap space $111$ and determined minimal patterns in the state transition graph that can correspond to an attractor that avoids $111$. We identified 6 patterns. In 5 cases of the 6, there exist edges whose linear extension eliminates the MAA, as shown in figure \ref{fig:prototypical_MAAs} in the Appendix. In Figure \ref{fig:3_prototypical_MAAs}  we illustrate the 2-variable system in which linear extension of any of the 4 edges eliminates the MAA (this system was introduced as Example \ref{ex:2d_MAA}), the 3-variable system in which each delay to a single self-edge eliminates the MAA but delays to other edges do not (this system was introduced as Example \ref{ex:template2iv}), and the 3-variable system that needs 2 delays to eliminate the MAA.

\begin{figure}
    \centering
    \includegraphics[width=1\linewidth]{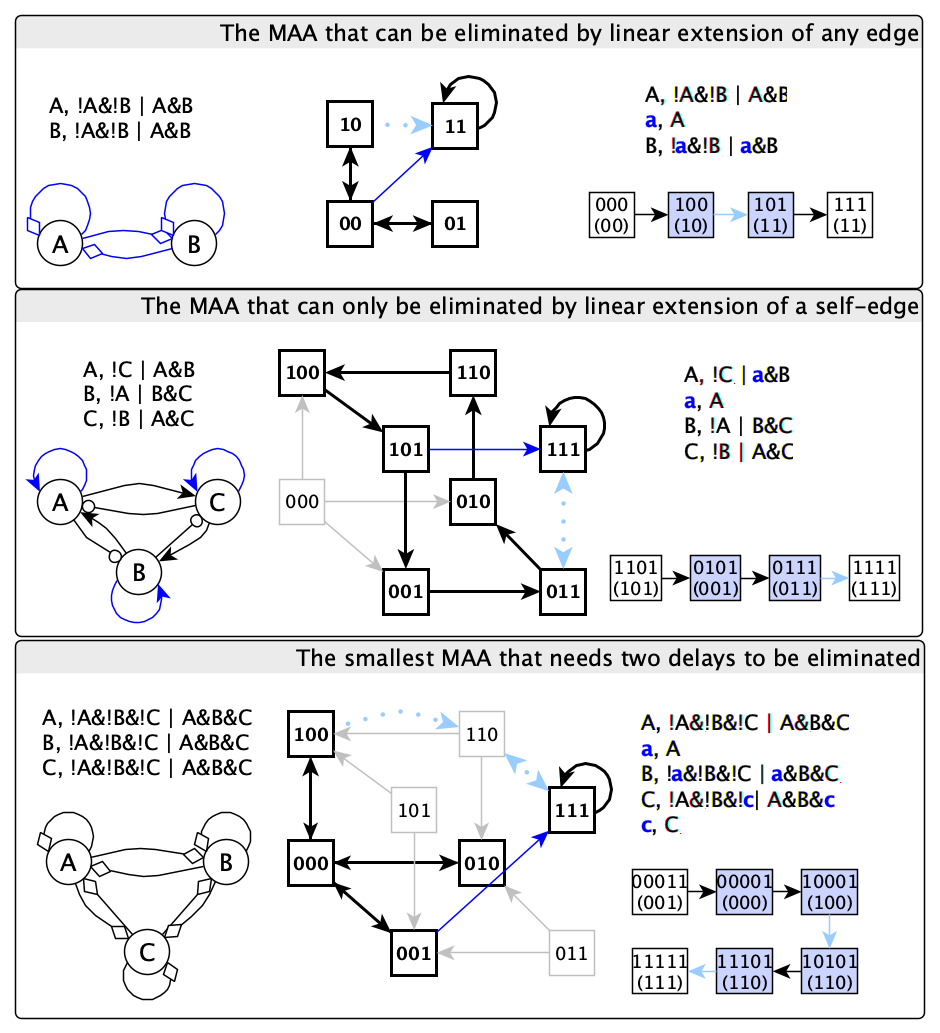}
    \caption{Three representative MAAs with minimal connectivity (i.e. eliminating any state transition among states of the MAA would disrupt it) and a trap space $111$. In the interaction graph a terminal arrow indicates a positive edge, a terminal circle indicates a negative edge and a terminal diamond indicates a dual edge. For the two MAAs that are possible to eliminate by a single linear extension, the corresponding edges are shown in blue in the interaction graph. For each MAA we indicate the projected STG for one example of a delay set that eliminates the MAA.}
    \label{fig:3_prototypical_MAAs}
\end{figure}

We found that all 9 MAAs that come from reduced versions of BBM-202 and BBM-151 can be eliminated by single delays. Figure \ref{fig:model_MAAs} depicts two such examples of reduced models.

\subsection{Motif-avoidant attractors in small RBNs are fragile to delays}

\begin{figure}
    \centering
    \includegraphics[width=0.8\linewidth]{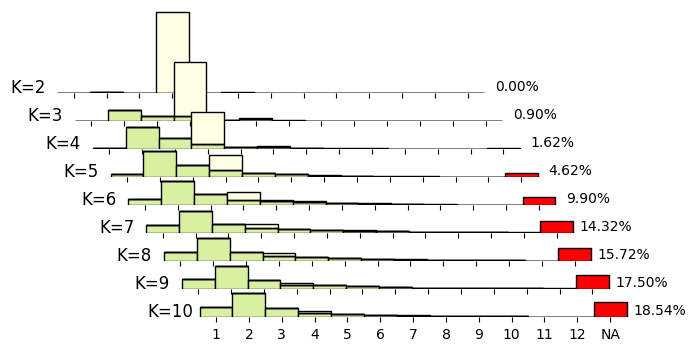}
    
    \caption{Histogram of the number of interactions among variables oscillating in an MAA wherein inserting a single delay (linear extension) destroys the motif avoidance for critical NK-networks with $N=10$ and $K$ between 2 and 10. Yellow bars represent MAAs that incorporate the two-variable MAA pattern shown in Example \ref{ex:2d_MAA}. Green bars are all other MAAs. Red bars and the percentages written next to them indicate cases where no single delay destroys the MAA.}
    \label{fig:single_delay_effect}
\end{figure}

To better understand the effect single-node delays have on motif avoidance, we return to the critical Random Boolean Networks, and select the $N=10$ ensemble across the full range of $K$ values. For each motif-avoidant attractor that was discovered in these networks, we compute the variables that oscillate in this attractor. We then introduce a delay node to each incoming edge of these oscillating variables (each delay is considered separately) and test whether motif avoidance was eliminated.\footnote{In the presence of multiple MAAs in one network, it is not clear how to associate the attractors of the original and the extended systems. In particular, even if the subspace of the original MAA no longer contains an attractor, a new MAA may appear somewhere else in the network. In such cases, we conservatively consider the delay to be ``working'' only if it eliminates all MAAs.}

The results of this analysis are presented in Figure~\ref{fig:single_delay_effect}, showing how the number of effective single-delay extensions changes with $K$. The percentage of MAAs that cannot be destroyed by a single delay is very small for $K\leq 5$ and even for $K=10$ it is less than $20\%$. We observe that for $K \leq 4$, the majority of MAAs can be destroyed by one of four linear extensions (for $K=2$ and $K=3$, this is almost all MAAs). We suspect that this is due to the two-variable MAA of Example~\ref{ex:2d_MAA}, which is indeed eliminated by extension of any of its four interactions (see also Figure~\ref{fig:3_prototypical_MAAs}, middle panel). In sparse networks, it seems unlikely that an MAA appears randomly due to some other combination of interactions. To test this, we further examine the state transition graph of each MAA, detect whether the four-state sub-graph corresponding to the two-variable MAA can be embedded into the detected MAA, and mark these cases with a lighter color in Figure~\ref{fig:single_delay_effect}.

For $K=4$ a secondary peak appears, indicating MAAs that can be eliminated by two separate single linear extensions. The simplest example of such MAA is a 3-variable system in which a 2-dimensional MAA of Example \ref{ex:2d_MAA} is inside the subspace in which the third variable is fixed in the state 0. We describe this example below and show it in Figure \ref{fig:example_figures} in the Appendix.

\begin{example} [Motif-avoidant attractor destroyable by two alternative single delays] \label{ex:two_single_delays} The system below has a trap space $111$ and a 2-dimensional motif-avoidant attractor inside the $**0$ subspace. 
\begin{bnexample}
\begin{BVerbatim}
A, (!A & !B & !C) | (A & B & C)
B,  (A & !B & !C) | (A & B & C)
C,  A & B 
\end{BVerbatim}
\end{bnexample}
Linear extension of the edge from $\node{A}$ to $\node{B}$ or of the edge from $\node{B}$ to $\node{A}$ eliminates the motif-avoidant attractor by creating a path to the canonical state $110$, which can reach the trap space. 
\end{example}

In conclusion, we observed the fragility of motif-avoidant attractors derived from random networks to linear extensions of single edges. MAAs of sparse networks are especially vulnerable, largely due to the over-representation of the 2-variable MAA of Example ~\ref{ex:2d_MAA}, which possesses maximal fragility, as extension of any of its edges eliminates motif avoidance. Positioning this motif-avoidant attractor inside a subspace that does not contain the trap space (Example~\ref{ex:two_single_delays}) reduces, but does not eliminate, its fragility.

\subsection{Efforts to identify an upper bound for the number of delays needed to eliminate a motif-avoidant attractor}

Because L-cuttability is sufficient for the non-existence of motif-avoidant attractors, it follows that adding delay nodes to all edges in an interaction graph will eliminate any MAA. Our analyses, however, have shown that in practice many fewer delays are needed. We therefore sought to lower the upper bound on the number of delays required to eliminate MAAs by leveraging dynamical information. We identified the size of the subspace that contains the MAA, the Hamming distance between the MAA and the avoided trap space, and the number of variables that have the same function as contributors to this upper bound. We started with simple and plausible conjectures (e.g., that the number of delays is bounded by the Hamming distance between the MAA and the trap space), but found counterexamples for all the conjectures we formulated. Finally, we identified an algorithm that is guaranteed to construct a path of state transitions from a state in the MAA to the trap space using delays on the regulations that drive the system toward the trap space. This algorithm yields an upper bound on the number of delays:  $(N-d)(N-1) + m(m+1)/2$, where $N$ is the number of variables, $d\leq N$ is the number of nodes that oscillate in the MAA, and $m<d$ represents the minimal disagreement between a state of the MAA with (the fixed nodes of) the trap space, when considering only the oscillating nodes of the MAA. In the worst case, with $m=d-1$ and a complete interaction graph with $E_{max}=N^2$ edges, the upper limit is $E_{max}-(\sqrt{E_{max}}-d/2)(d+1)$, which is strictly smaller than $E_{max}$.

The validity of the algorithm automatically proves that the aforementioned upper bound is true.
The algorithm utilizes the fact that the subspace of the MAA is not a trap space, and hence, there exists an escape transition out of it. We align a node of the MAA with the trap space by making this node have access to the escape transition. This is done by adding delays on the incoming edges of the node from the oscillating nodes of the MAA, allowing the node's update function to read the needed state of the subspace.
We indicate the full description of the algorithm and derivation of the number of delays in Appendix \ref{sec:proof_of_upper_bound}.  Figure \ref{fig:star_MAA_elimination} in the Appendix presents an example of applying the algorithm to a system with $N=4$, $d=3$ and $m=1$. 
 
If $N=d$ and $m=1$, the upper bound is 1, meaning that the motif-avoidant attractor can be eliminated by a single linear extension. Application of the algorithm indicates that this linear extension is to the self-regulation of the node whose state differs in the MAA compared to the trap space. Six of the seven prototypical 3-variable MAAs shown in Figure \ref{fig:prototypical_MAAs} have $N=d$ and $m=1$ and indeed can be eliminated by a delay to a single self-regulation.

However, many of our example MAAs and model-derived MAAs illustrate ways in which the upper bound overestimates the needed number of delays (see Figure \ref{fig:model_MAAs} and Figure \ref{fig:example_figures} in the Appendix). 
The algorithm above relies on the regulations that are part of the stable motif (which lead to the self-transitions of the states of the trap space). In case of a large Hamming distance between the motif-avoidant attractor and the trap space, using delays to edges that involve states within the MAA offers a potentially better alternative. Indeed, we have identified a family of MAAs that can utilize this aspect. The number of delays necessary and sufficient to eliminate this MAA family, ($\sim E_{max}/4$), is lower than the upper bound. Hence this number represents a lower bound of the number of delays needed in a general algorithm that can eliminate any MAA.

\begin{example} [The number of delays necessary and sufficient to eliminate this motif-avoidant attractor family scales linearly with $E_{max}$] \label{ex:star_graph_family}
\begin{bnexample}
\begin{equation*}    
    \texttt{X}_i,\texttt{ !X}_1\texttt{ \& }\texttt{!X}_2\texttt{ \&...\& }\texttt{!X}_N\texttt{ | }\texttt{X}_1\texttt{ \& }\texttt{X}_2\texttt{ \&...\& }\texttt{X}_N
\end{equation*}
\end{bnexample}
\end{example}
This family of systems has a point attractor $11..1$ and a motif-avoidant attractor with $N+1$ states, including $00..0$ and all states in which one variable is 1 and the rest are 0. Because of the star-like shape of the motif-avoidant attractor, we will refer to this family of systems as the star-shaped MAA. Note that we will always consider this MAA in combination with the trap space $11..1$. The interaction network of this system is complete, with $E_{max}=N^2$ edges. 

According to the upper bound found above, the elimination of this star-shaped MAA should require $N(N-1)/2$ delays. However, the $N=3$ member of this family needs delays to only two edges (instead of 3) to be destroyed. Figure \ref{fig:3_prototypical_MAAs} (bottom row) illustrates one possible delay combination. The smaller number of delays is made possible by utilizing a transition within the MAA, in this case the transition from $000$ to $010$. 

Similarly, the $N=4$ member of the family requires 4 delays instead of 6 (compare the second and third panels of  Figure \ref{fig:star_MAA_elimination} in the Appendix). We determined that in the general case the number of delays necessary and sufficient for the elimination of the MAA is $N^2/4$ for even $N$ and $(N^2-1)/4$ for odd $N$ (see Appendix \ref{sec:proof_of_star_shaped_MAA_elimination} for the derivation). Thus, this example has a quadratically scaling number of necessary delays, as the upper bound, but the number of delays is less than (about half of) the upper bound. 

In general, having the choice of adding delays such that a transition within the MAA (as an alternative to a transition within the trap space) is utilized to align a variable with the trap space is expected to reduce the number of delays needed to eliminate the MAA. In both cases the transition starts from a memory state of the delay-expanded system. In order for this transition to eliminate the MAA, this memory state needs to be reachable from the canonical states derived from the original system's MAA. We discovered that the general reachability of memory states is not guaranteed. The reason for the loss of reachability is the loss of transitions in certain memory states compared to the original system (and to canonical states). When a memory state affected by such loss becomes a Garden of Eden state, it and potentially additional memory states become unreachable from the MAA. In Appendix \ref{sec:unreachability_issue} we indicate an example of a delay failing to eliminate the MAA of Example \ref{ex:template2iv} due to an unreachable memory state and contrast it to the effectiveness of a delay affecting a self-regulation. Our algorithm for the upper bound of the number of delays is designed such that it utilizes only the reachable memory states. The star-shaped MAA family avoids unreachability issues due to their unique shape. While the unreachability issue may not be avoidable in general, we did not find any examples in which it caused the lack of all alternatives to destroy an MAA. We indicate more details in  Appendix \ref{sec:unreachability_issue}.

We finish this section by turning back from worst-case scenarios of MAA elimination to empirically observed elimination of MAAs.

\subsection{The motif-avoidant attractors in reduced real-world models are non-biological and easily eliminated or trapped}\label{sec:maas_in_published_models}

Among the 230 BBM models, we found two models whose reduced versions exhibited motif-avoidant attractors. The first model \cite{gupta2022boolean} describes signal transduction in non-small cell lung cancer. It has 28 nodes, including the input node ``$\node{DNA-damage}$'', and 107 edges. The original article focuses on the setting $\node{DNA-damage}=1$ and reports two point attractors, describing cell cycle arrest and apoptosis, respectively. The setting $\node{DNA-damage}=0$ corresponds to cancer cell proliferation. The MAA only appears in this latter setting, in reduced versions of the model with 4 to 8 nodes. Its appearance is enabled by an unintended feature of the update function of $\node{p53-A}$, a modified form of the tumor suppressor protein $\node{p53}$, which allows the activity of $\node{p53-A}$ even if the parent $\node{p53}$ protein is inactive. This outcome of the function is inconsequential in the setting of $\node{DNA-damage}=1$, thus it likely avoided the attention of the modelers. 

\begin{figure}
    \centering
    \includegraphics[width=1\linewidth]{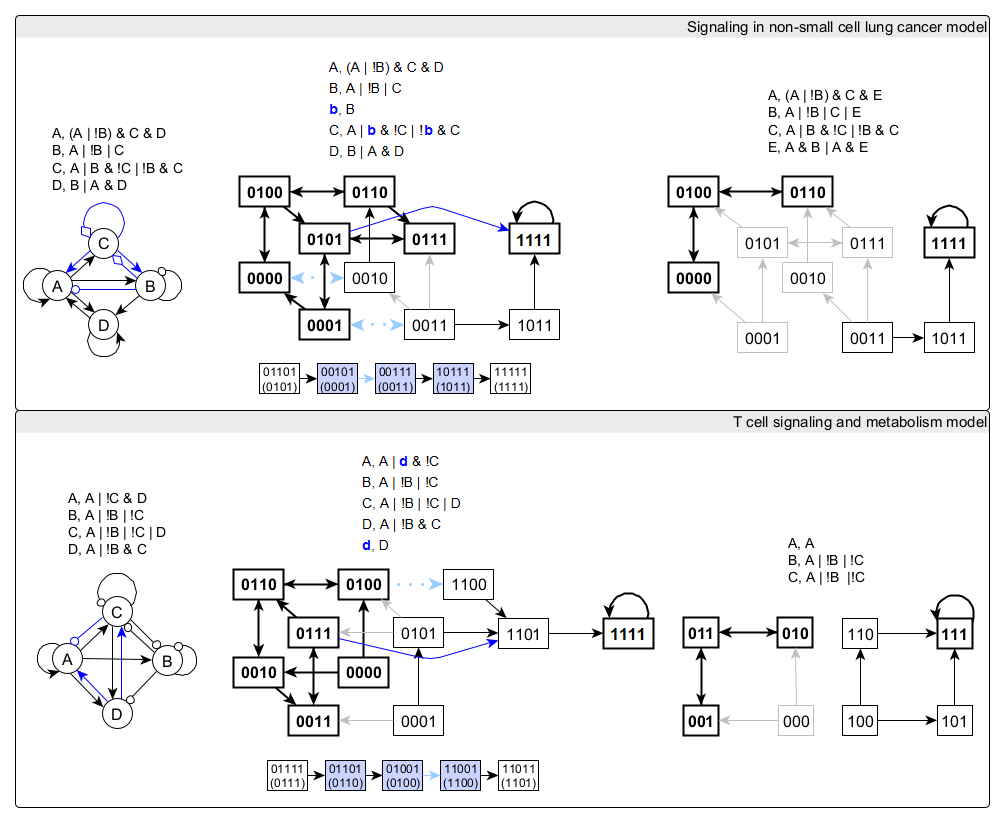}
    \caption{Two motif-avoidant attractors that come from reduced versions of Boolean models. For each system, we indicate the interaction graph and the Boolean functions. Multiple alternative single delays can eliminate both systems' motif-avoidant attractors. We indicate the edges whose individual linear extension can eliminate the MAA in blue and we give the projected state transition graph corresponding to one such example for each network. Both model-derived MAAs illustrate that escaping the subspace of the MAA can be accomplished by linear extension of a subset of the edges incident on the nodes fixed by the MAA. For visual clarity, the projected STGs show 10 or 11 states instead of 16. For each system we also include an overlapping or reduced version that has an oscillation contained in a minimal trap space instead of a motif-avoidant attractor.}
    \label{fig:model_MAAs}
\end{figure}

We illustrate a four-node reduced version of the model in Figure \ref{fig:model_MAAs}. For visual clarity, we shortened the variable names and performed a parity change in three of the variables. Node $\node{A}$ of this system corresponds to the activity of the protein $\node{c-Myc}$ in the original model, node $\node{B}$ represents the inactivity of $\node{p53}$, node $\node{C}$ represents the inactivity of $\node{p53-A}$, and node $\node{D}$ represents the inactivity of the microRNA $\node{miR-34a}$. The system has a trap space $1*11$, a minimal trap space (point attractor) $1111$, as well as an MAA that visits 6 states of the $0***$ subspace. The point attractor agrees with a proliferative phenotype, while the MAA cannot be attributed biological meaning. The state $0011$ has a transition to $1011$, thus the $0***$ subspace is not a trap space. The MAA can be eliminated by a linear extension of a single edge; the five edges that are successful candidates are indicated by blue color in the interaction graph. We also provide the projected STG for one of these extensions. Alternative reductions of the model feature an attractor that is contained inside a minimal trap space. We illustrate one such model, which overlaps the previous example in three nodes and has the additional node $\node{E}$, which represents the cell cycle protein $\node{YY1}$. This system has a minimal trap space $1111$ as well as a 3-state oscillation within the $0**0$  trap space. 

The second model~\cite{sanchez2019contribution} connects metabolism, the production of reactive oxygen species ($\node{ROS}$), and T cell signaling. It contains 111 nodes, including two input nodes, the T cell receptor $\node{TCR}$ and the co-stimulatory protein $\node{CD28}$, and 244 edges. In the original model 21 nodes have 3 states (0, 1, 2), with separate functions for the level 1 and the level 2. The model was made Boolean by representing the two levels as two separate nodes. The original study reports that in the absence of $\node{TCR}$ or $\node{CD28}$, cells remain quiescent or are kept in metabolic anergy (inactivity). In the presence of both $\node{TCR}$ and $\node{CD28}$ signals, cells could either adopt an anergic state with high $\node{ROS}$, or undergo activation. However, this model includes a hidden constant that is introduced by the Booleanization process: the function that describes level 1 of node $\node{ETC}$, which represents the enzymes involved in the electron transport chain, always evaluates to 1. Our method interpreted this constant as a source node, and found MAAs in the setting where this constant level is set to 0. Biologically, this corresponds to a perturbation of the original system that blocks the activity of the electron transport chain, coupled with the inactivity of either $\node{TCR}$ or $\node{CD28}$. In other words, the conditions captured by the MAA do not represent a biological setting originally considered when designing the model, but rather a perturbation whose biological realizability has not been tested.

We illustrate a four-node reduced version of the model in Figure \ref{fig:model_MAAs}. Node $\node{A}$ represents a high level of mitochondrial $\node{ROS}$, node $\node{B}$ represents the inactivity or the ATP synthase enzyme, node $\node{C}$ represents a lack of thioredoxin pool in the mitochondria, and node $\node{D}$ represents a high activity of the enzyme thioredoxin reductase. This system has a stable motif $\node{A}=1$, a minimal trap space $1111$, as well as an MAA that visits 5 states of the $0***$ subspace. The point attractor corresponds to an anergic state with high $\node{ROS}$, while the MAA does not have a known biological interpretation. The motif-avoidant attractor can be eliminated by a single delay to any of three edges (shown in blue in the interaction graph).  This model can be simplified further by reducing the node $\node{D}$. The 3-node system has a minimal trap space $111$ and instead of the MAA has a 3-state oscillation contained in the minimal trap space $0**$. The reason for this change in the nature of the oscillation is the emergence of a new stable motif, $\node{A}=0$. This stable motif arises because upon substituting the function of $\node{D}$ into the function of $\node{A}$, the second clause becomes "\texttt{!C \& !B \& C}", which evaluates to 0.

In summary, the motif-avoidant attractors in both models are unintended and appear only for specific input values and specific instances of model compression. Other choices of model compression lead to trapped oscillations or no oscillatory attractor. The MAAs are prone to elimination by single linear extension and prone to trapping by node reduction.

\section{Discussion}

Motif avoidance plays an important role in synchronous update and underlies the cases in which a sustained oscillation arises from a positive feedback loop (as opposed to a negative feedback loop). Synchronous MAAs are a significant contributor to the substantially higher number of attractors in Random Boolean Networks (RBNs) under synchronous update \cite{kaufman2005scaling} compared to asynchronous update \cite{greil2005dynamics, rozum2021parity}.  Multiple Boolean models of biological systems that yield sustained oscillations rely on motif avoidance under synchronous update~\cite{park2023models, dahlhaus2016boolean, orlando2008global}. Yet, the frequency of motif-avoidant attractors under asynchronous update has not been systematically studied before.

To fill this knowledge gap, we performed an extensive search in 230 published Boolean models with 14~136 input combinations and more than 100 million RBNs. We found that MAAs are very rare. There were no MAAs in the published Boolean models.  A small percentage of small and dense RBNs have motif-avoidant attractors ($\sim1.6\%$ for complete networks of 10 nodes) and in large, sparse RBNs the MAA likelihood is astronomically small. Nevertheless, in cases where it is important to exactly enumerate attractors, it is both critical and computationally difficult to account for MAAs~\cite{rozum2021parity}.

Moreover, MAAs can be created (or destroyed) by node deletion reduction. Our numerical analysis of reduced versions of the Boolean models and maximally reduced versions of RBNs indicates that network reduction overall increases the MAA likelihood. The MAA likelihood in maximally reduced RBNs is similar to the MAA likelihood of small and dense networks. Furthermore, we identified MAAs in reduced versions of two published Boolean models. We could not ascribe biological relevance to either MAA. Our results suggest that compression of information during model construction may inadvertently lead to the emergence of MAAs via surprising non-local network effects.

By analyzing our ensemble of MAAs arising in RBNs, we found that they are fragile to linear extension of single edges. We have proved that an MAA in the same subspace as an avoided trap space can always be eliminated by a single linear extension if the minimal Hamming distance between the MAA and the trap space is one. This situation occurs frequently in small, dense Boolean networks. Indeed, we found that the majority of MAAs represented in our RBN ensembles are the smallest and most fragile MAAs, presented in Example \ref{ex:2d_MAA}. The MAAs that arise from reduced versions of Boolean models of biological systems, though they are more complex (they have more variables and are not in the same subspace as the avoided trap space), can also be eliminated by single linear extensions. These results pose an interesting contrast to the currently known upper bound for the number of delays needed to eliminate a motif-avoidant attractor, namely, that linear extensions of up to all the edges are needed such that the interaction graph becomes L-cuttable. We lowered the upper bound using limited dynamic information, and showed that this bound cannot be lowered below $\floor{N^2/4}$.

The currently known theoretical criteria for ruling out MAAs based on the interaction network \cite{naldi2023linear, richard2023attractor} are strict. We noted that only 9 of the 230 published Boolean models have an L-cuttable interaction graph. Yet, we also noted that when considering fixed values of the input nodes (which denote external signals or cellular context), and percolating these values through the model, $38\%$ of the instances yield an empty network (indicative of monostability for each input configuration), and $9\%$ of the instances yield an L-cuttable network. This simplification of the `effective' interaction graph based on dynamic context may contribute to the lack of MAAs in biologial networks and warrants further study, for example using the methods of~\cite{gatesEffectiveGraphReveals2021}.

Significant open questions remain. For example, at present, there is no estimate of the prevalence of the unreachability issue (Appendix~\ref{sec:unreachability_issue}) that limits the candidates for linear extensions to regulations of the trap space. We discuss potential leads in Appendix \ref{sec:unreachability_issue}. We expect that the upper bound for the number of delays necessary to eliminate an MAA can be lowered in the classes of motif-avoidant attractors in which this issue can be ruled out; this may include all MAAs.

The search for simple criteria for the presence or absence of motif avoidance should continue. From previous theoretical work~\cite{richard2007necessary,richard2010negative, richard2023attractor, naldi2023linear} we know that Boolean networks whose interaction graph has no cycles, has only positive or only negative cycles, lacks a path from a positive cycle to a negative cycle, or is L-cuttable cannot have MAAs. These criteria leave many possibilities, especially in dense networks, in which motif avoidance cannot be ruled out. We considered several natural candidates for additional criteria on the structure of the interaction network, or on the Boolean update functions and found counterexamples for them. For instance, although most systems that have asynchronous MAAs have non-monotonic update functions, Example \ref{ex:monotonic_nested} shows that nested canalizing functions, thought to be most biologically relevant, can also lead to motif avoidance. Although many systems with MAAs have negative or ambiguous self-edges, Example \ref{ex:template1_delay} shows that motif avoidance is possible in systems that lack self-regulation. 

Increased clarity on the minimal number of delays to eliminate a motif-avoidant attractor will help with the biological interpretation of motif avoidance. MAAs under asynchronous or random set update retain an element of synchronicity: the effect of a variable's state change is immediately reflected in all update functions it regulates. Though this may be a useful approximation in many cases, it fails to account for inherent delays in a system, such as when a protein must slowly diffuse through the cytoplasm to reach its target. Boolean systems with explicit delays associated with each edge are a way of including such non-negligible durations~\cite{cheng2013autonomous, akman2023bdetools}. Identifying which delays affect the long-term dynamics of a Boolean network would be useful in building more biologically realistic models. 

A clearer picture of the reasons that motif-avoidant attractors are rare in asynchronous Boolean networks is beginning to emerge. One hint is that the the small and dense subnetworks associated with MAAs are rare in biologically relevant Boolean models (whether empirical or random). Another hint is the strong canalizing role played by input nodes in Boolean models of biological systems, which fix large portions of the network, making many network edges functionally redundant. This effectively eliminates any dense subnetworks that do appear and might otherwise give rise to motif avoidance. A third hint is the fragility of MAAs to single delays, which we have shown can arise when MAAs closely approach avoided trap spaces. These hints do not yet give the whole picture, but they may point the way to a more complete understanding and computationally efficient characterization of motif avoidance.

\section*{Funding}

Ond\v{r}ej Huvar has been supported by the Czech Science Foundation grant No. GA22-10845S. Samuel Pastva received funding from the European Union’s Horizon 2020 research and innovation programme under the Marie Sklodowska-Curie Grant Agreement No. 101034413. Kyu Hyong Park and R\'eka Albert have been supported by NSF grant MCB 1715826 and ARO grant 79961-SM-MUR. No funding bodies had any role in study design, analysis, decision to publish, or preparation of the manuscript.

\section*{Data availability}

The networks, code and raw results upon which we base our analysis are available as part of a reproducibility artefact at \url{https://doi.org/10.5281/zenodo.13860057}.

\bibliography{sn-bibliography}

\begin{appendices}
\section{Supplementary information on the number of delays needed to eliminate a motif-avoidant attractor}
\label{sec:supplementary_text}

\subsection{Proof of the validity of the algorithm and upper bound for the number of delays needed to eliminate a motif-avoidant attractor}
\label{sec:proof_of_upper_bound}

Consider a Boolean network of $N$ nodes that contains a known stable motif and its trap space $\mathcal{X}$ in which $n$ variables are fixed (and $N-n$ are free to vary), and a $d$-dimensional MAA $\mathcal{A}$. Without loss of generality, we take the first $n$ variables of the network to be in the ON state within $\mathcal{X}$, and also assume that the smallest trap space that contains both $\mathcal{A}$ and $\mathcal{X}$ is the whole state space. In the following we describe an algorithm that can be used to determine the set of delays (linear extensions of edges) which if simultaneously added to the system will eliminate the MAA. For simplicity, in the following description we will refer to states (not necessarily unique) of the delay-expanded system by indicating the states of the original (non-delay) nodes. The corresponding states in the delay-expanded system are canonical states, or memory states in which each delay node acts as a memory of a past state of its parent node.

First, we show that in the worst case, $(N - d)(N - 1)$ delays are sufficient to expand the MAA such that it oscillates in all $N$ variables. Consider the $N-d$ nodes $\node{F}_i$ that are fixed to state $f_i$ in $\mathcal{A}$. By adding delays to all the incoming edges of these nodes (except self-edges), we ensure that all variables $\node{F}_i$ can read any state in the $\node{F}_i=f_i$ subspace.

This can be shown in an iterative way. Since the subspace that contains $\mathcal{A}$ (which has $d$ oscillating variables) is not a trap space, one of its states must have a transition that turns a variable among $\node{F}_i$ to its opposite state, say variable $\node{F}_1$ to state $\bneg f_1$. Having delay nodes on all incoming edges of $\node{F}_1$ allows it to read this state and make the transition. In other words, as the delayed system oscillates through the states that correspond to $\mathcal{A}$ (but with new delay nodes), the $d$ variables oscillate, and the delay nodes can retain any combination of these values. This is possible because the delay nodes only affect $\node{F}_1$, and the transitions within $\mathcal{A}$ do not involve the update of $\node{F}_1$. Thus it is possible for the delay nodes to be updated to a combination that matches the state where $\node{F}_1$ can turn to state $\bneg f_1$. Note that this allows $\node{F}_1$ to turn to state $\bneg f_1$ at any state that corresponds to $\mathcal{A}$ (but with the aforementioned delay node states fixed to the required values). If $\node{F}_1=\bneg f_1$ creates a path to the previously avoided trap space, the goal of eliminating the MAA is achieved. Otherwise, the transition will lead back to the MAA, and $\node{F}_1$ will oscillate. Let's call the resultant attractor $\mathcal{A}^{(1)}$. 

Next, consider a larger subspace (of the original system) wherein $\node{F}_1$ and the $d$ variables are free to vary. Again, since this subspace is not a trap space, there must exist a state in which a variable $\node{F}_2$ turns to state $\bneg f_2$. Adding delays to all the incoming edges of $\node{F}_2$ allows this transition from any state that corresponds to $\mathcal{A}^{(1)}$, again because the $d$ nodes and $\node{F}_1$ oscillate in $\mathcal{A}^{(1)}$, and the new delays only affect $\node{F}_2$. Similarly, when all $d$ variables and the originally fixed variables up to $\node{F}_k$ are allowed to oscillate in $\mathcal{A}^{(k)}$, having delays on all incoming edges for variable $\node{F}_{k+1}$ allows it to turn to state $\bneg f_{k+1}$ from any state that corresponds to $\mathcal{A}^{(k)}$. By induction, having delay nodes on all the incoming edges of $\node{F}_i$ is sufficient to guarantee the transition of all $\node{F}_i$ variables to the $\bneg f_i$ state. In the worst case, each node $\node{F}_i$ has $N-1$ regulators other than itself, leading to an upper bound of $(N-d)(N-1)$ delays added in this process. Note that at this point, all $N$ original variables oscillate in the MAA $\mathcal{A}^{(n)}$. Furthermore, from any state that corresponds to $\mathcal{A}$, we can make all the $\node{F}_i$ variables to agree with $\mathcal{X}$ in a sequential manner. However, this still does not guarantee that the avoided trap space can be reached.

Next, consider the $d$ nodes $\node{A}_i$ that initially oscillate in $\mathcal{A}$. Among the states of $\mathcal{A}$, consider a state $S$ wherein the most nodes among $\node{A}_i$ agree with $\mathcal{X}$. As described above, using at most $(N-d)(N-1)$ delays on the edges incident on $\node{F}_i$, we can change the state of all the variables $\node{F}_i$ to agree with $\mathcal{X}$. Let's say this is done by following a path of state transitions $\mathcal{P}$. $\mathcal{P}$ visits states in which each of the $d$  nodes are ON, states in which each of the $d$ nodes are OFF, lead to state $S$, turn variables $\node{F}_i$ ON, resulting in state $S_{n-m}$. $S_{n-m}$ has $m$ nodes among $\node{A}_i$ that disagree with $\mathcal{X}$ (i.e., they are in the OFF state), while all nodes $\node{F}_i$ agree with $\mathcal{X}$ (i.e., they are in the ON state). Such a path always exists because every node among the $d$ nodes $\node{A}_i$ visits its ON and OFF state within the attractor, and the delay nodes incident on each $F_i$ have access to both the ON and OFF state of all of the $d$ nodes and $F_{j<i}$. In our later steps, we omit referring to the $(N-d)(N-1)$ delay nodes for simplicity; they are implicitly updated appropriately along $\mathcal{P}$.

Therefore the task is to turn the $m$ nodes $\{\node{X}_i\}$ ON in $S_{n-m}$. Here the set of nodes $\{\node{X}_i\}$ denotes the subset of nodes of $\node{A}_i$ that needs to be updated in $S_{n-m}$ to reach the avoided trap space. Note that due to the self-transitions of $\mathcal{X}$, each node $\node{X}_i$ can turn ON if all other nodes among the $n$ nodes are ON. We utilize this by accessing the history of ON states of these nodes previously visited within the attractor.

When the path $\mathcal{P}$ visits a state with node $\node{X}_j$ ON, the $\node{X}_j=\text{ON}$ state can be stored in a delay node; consider that this delay node affects some other node $\node{X}_i$. Our aim is to make sure that each node $\node{X}_i$ can have access to the ON states of the delays of its regulators in state $S_{n-m}$, so that they can turn ON. Note that since $\node{X}_i$ is OFF in $S_{n-m}$, the path must turn OFF $\node{X}_i$ at least once. However, $\node{X}_i$ may be unable to turn OFF if the state of the incident delay node is ON. (If the update function of $\node{X}_i$ requires the OFF state of $\node{X}_j$ for the OFF state of $\node{X}_i$, the only way to accomplish this in the delayed system is to have the delay node OFF.)

This problem can be avoided if we arrange $\{\node{X}_i\}$ in the order of their last ON states, i.e., so that $\node{X}_1$ is the node whose last ON state is the first among  $\node{X}_i$ in the path, $\node{X}_2$ is the node whose last ON state is the second, and $\node{X}_m$ is the node whose last ON state is the last in the path. Note that each of the $\node{X}_i$ nodes may turn ON multiple times, but we only consider their last ON state. This allows the delay nodes of $\{\node{X}_{j> i}\}$ to be updated to ON \emph{after} $\node{X}_i$ turns OFF. In the case of a self-edge, a delay of $\node{X}_i$ ON does not prevent $\node{X}_i$ from turning OFF. (Since the delay node and the parent node is in the same state, if $\node{X}_i$ could turn OFF in the original system, it can still turn OFF in the delayed system.)

Let $\node{X}_1$ have delays on its edges that start from the rest of the nodes $\{\node{X}_i\}$ (including itself if it has a self-edge, at most $m$). Consider an instance of $\mathcal{P}$, $\mathcal{P}^{(1)}$, which visits a state where $\node{X}_1$ is ON, visits states where $\node{X}_i$ is ON in a sequence without turning $\node{X}_1$ ON, and reaches $S_{n-m}^{(1)}$, while updating the delay nodes incident on $\node{X}_{1}$ ON. At $S_{n-m}^{(1)}$, all the delay nodes incident on $\node{X}_1$, along with the other $n-m$ nodes, are ON. This allows $\node{X}_{1}$ to turn ON using the self-transitions in $\mathcal{X}$ (which make the stable motif "stable"). This leads to a state $S_{n-(m-1)}^{(1)}$ that has $m-1$ nodes among $\node{A}_i$ that disagree with $\mathcal{X}$. We repeat the process by letting $\node{X}_{2}$ have delays on its edges that start from the nodes $\node{X}_{i\geq 2}$ (at most $m-1$). A new path $\mathcal{P}^{(2)}$ again visits states where $\node{X}_1$, $\node{X}_2$, and the rest of $\node{X}_i$ is ON in a sequence, just as in $\mathcal{P}^{(1)}$, but it also updates the delay nodes incident on $\node{X}_{2}$. At $S_{n-m}^{(2)}$, $\node{X}_{1}$ can be updated to ON same as before to reach $S_{n-(m-1)}^{(2)}$. Here the delay nodes incident on $\node{X}_{2}$, $\node{X}_1$ itself, and the other $n-m$ are in their ON states. This allows $\node{X}_{2 }$ to turn ON and lead to state $S_{n-(m-2)}^{(2)}$. Repeating this process leads to $S_{n}^{(m)}$, which is part of $\mathcal{X}$, using at most $m(m+1)/2$ delays.

As a result, having $(N-d)(N-1) + m(m+1)/2$ delays guarantees the elimination of the MAA. Here $m$ represents the minimal disagreement between a state of the MAA with (the fixed nodes of) the trap space, when this disagreement is considered only among the nodes $A_i$.

The first panel of Figure \ref{fig:star_MAA_elimination} illustrates the application of the algorithm to eliminate an MAA with $N=4$, $d=n=3$ and $m=1$. Node $\node{D}$ is in the OFF state in the four states of the MAA. The transition $(1~1~1~0) \rightarrow (1~1~1~1)$ is an escape from the $***0$ subspace. Adding delay nodes to the edges from $\node{A}$, $\node{B}$, and $\node{C}$ to $\node{D}$ adds 7 new transitions from memory states with $D=0$ and the ON state of the delay nodes to memory states with $D=1$. These transitions expand the MAA from $4$ states with $D=0$ to $8$ states with oscillating $\node{D}$ when considering the states of the original system. (When considering all delay nodes, the expansion is to $52$ states. This includes all the $32$ states derived from the 4 states of the original MAA, as well as $20$ additional states with $\node{D}$ in the ON state.) As there isn't a path from the expanded MAA to the trap space, more delays are needed. The state of the expanded MAA with fewest disagreements with the trap space is $(01~01~11~1)$ (or $0011$ considering only the original nodes), in which the state of node $\node{B}$ disagrees with the trap space. Adding a delay on the self-regulation of $\node{B}$ allows it to turn ON from the memory state $(01~011~11~1)$, which is part of the strongly connected component of the delay-expanded system's STG. The transition creates a path to the trap space, thus there no longer is an MAA.

\subsection{Derivation of the number of delays needed to eliminate the motif-avoidant attractor of Example \ref{ex:star_graph_family} }
\label{sec:proof_of_star_shaped_MAA_elimination}

Consider the family of motif-avoidant attractors defined in Example \ref{ex:star_graph_family}, to which we will refer to as the star-shaped MAA. Here we present the details of an algorithm to eliminate star-shaped MAAs using transitions both within the MAA and within the trap space. The number of delays indicated by this algorithm is one half of the upper bound obtained if only using the self-transitions within the trap space. We then prove that this number of delays cannot be reduced further.

For simplicity, in the following description we will refer to states (not necessarily unique) of the delay-expanded system by indicating the states of the original (non-delay) nodes. The corresponding states in the delay-expanded system are canonical states, or memory states in which one or more delay nodes act as a memory of a past state of their parent nodes. We will denote these states using as subscript the number of original nodes that are ON. In our proofs, we describe the addition of delays in an iterative way, but in practice they are all added at once. We distinguish the different iterations using superscripts. For example, state $S_{1}^{(1)}$ and state $S_{1}^{(2)}$ will have the same state of the original nodes, but $S_{1}^{(2)}$ is a state of the system expanded with more delay nodes.

\paragraph{Algorithm to eliminate the star-shaped MAA:} 

In the star-shaped MAA the state $00..0$ can turn ON any node $\node{X}_i$. Suppose we start from a state $S_1$ where node $\node{X}_1$ is ON (all other nodes OFF), and we need to turn the remaining $N-1$ nodes ON to reach the trap space $11..1$.

\textbf{Claim 1:} The star-shaped MAA can be destroyed by placing $N-k+1$ delay nodes incident on each $\node{X}_k$ ($1<k\leq N$). That is, $\frac{(N-1)N}{2}$ delays overall.

One can follow the algorithm for the upper bound of delays (specifically the second part, since all variables already oscillate in the star-shaped MAA) to accomplish this goal and verify the validity of the claim. The delay nodes incident on $\node{X}_{k}$ serve as a memory of the ON state of their parent nodes $\{\node{X}_{i\geq k}\}$, allowing $\node{X}_{k}$ to turn ON using a self-transition within the trap space.

\textbf{Claim 2:} The star-shaped MAA can be destroyed by placing $k-1$ delay nodes incident on each $\node{X}_{k}$ ($1<k\leq N$). That is, $\frac{(N-1)N}{2}$ delays overall.

This claim presents a symmetrical counterpart to Claim 1. In Claim 1 we added $N-1$ delays, one on each edge from nodes $\{\node{X}_{j \geq 2}\}$ to node $\node{X}_{2}$, to turn $\node{X}_{2}$ ON. Instead, we can achieve the same with a single delay, on the edge from $\node{X}_1$ to $\node{X}_{2}$. Consider a transition that starts from the state $S_{0}^{(2)}$ (which equals the canonical state $00..0$), turns $\node{X}_{1}$ ON, and reaches state $S_{1}^{(2)}$. Then the delay node is OFF in state $S_{1}^{(2)}$, which allows $\node{X}_{2}$ to turn ON, thereby reaching state $S_{2}^{(2)}$. Similarly, add two delays on the edges from the nodes $\node{X}_{1}$ and $\node{X}_2$ to $\node{X}_3$. As the system transitions from $S_{0}^{(3)}$ to $S_{2}^{(3)}$, the delay nodes preserve the memory of the OFF state of $\node{X}_{1}$ and $\node{X}_2$, allowing $\node{X}_3$ to turn ON at $S_{2}^{(3)}$ and reach $S_{3}^{(3)}$. Repeating this process, suppose there exists a path that starts at $S_{0}^{(k-1)}$, goes through $S_{1}^{(k-1)}$, $S_{2}^{(k-1)}$, ... and arrives at state $S_{k-1}^{(k-1)}$ that has $k-1$ original (non-delay) nodes, $\node{X}_{1}$ to $\node{X}_{k-1}$, ON. Adding $k-1$ delay nodes, on the edges from nodes $\{\node{X}_{j\leq k }\}$, allows the system expanded with the new delay nodes (which are kept in the OFF state) to follow a path almost identical to the previous; this path visits the states $S_{1}^{(k)}$, $S_{2}^{(k)}$, ... , $S_{k-1}^{(k)}$. Then $\node{X}_{k}$ can be updated to the ON state at state $S_{k-1}^{(k)}$, reaching $S_{k}^{(k)}$. By induction, this process can reach the state $S_{N}^{(N)}$ wherein all original nodes are ON, and then update the delay nodes to reach the trap space. In summary, the delay nodes serve as a memory of the OFF state of their parent nodes $\{\node{X}_{i<k}\}$, allowing $\node{X}_{k}$ to turn ON using a transition within the MAA.

\textbf{Claim 3:} These two delay schemes are compatible and therefore the star-shaped MAA can be destroyed by placing EITHER $k-1$ or $N-k+1$ delays on each $\node{X}_{k}$. (The minimal number of added delays in this scheme is characterized by Claim 4.)

In the system without delay nodes, $\node{X}_{k}$ can be updated to the ON state either if all its regulators are OFF, or if all its regulators are ON. In a system with delay nodes, $\node{X}_{k}$ can be updated to the ON state from the state $S_{k-1}$ if either the delay nodes in the edges from $\{\node{X}_{i\geq k}\}$ are in their ON states, or if the delay nodes in the edges from $\node{X}_{i<k}$ are in their OFF states. We show that there exists a path that allows either, independent of the choices for delays for other nodes  $\node{X}_{j}$.

Suppose that we have a delay-expanded system that has delays on the edges to nodes $\{\node{X}_{1<i<k}\}$, in such a way that there exists a path $\mathcal{P}^{(k-1)}$ from $S_0^{(k-1)}$ (where all original nodes are OFF) to $S_{k-1}^{(k-1)}$ (where the original nodes $\{\node{X}_{i<k}\}$ are ON). Note that in general $S_0^{(k-1)}$ is not a canonical state, and must have appropriate states of the delay nodes on the edges to nodes $\{\node{X}_{i<k}\}$ to allow $\mathcal{P}^{(k-1)}$. Also assume that there exists a path $\mathcal{P'}^{(k-1)}$ that starts at the canonical state $00..0$ in the MAA, visits states where each of the nodes (starting from $\node{X}_1$ and ending at $\node{X}_N$) is ON individually, while going back to a canonical or memory variant of $00..0$ each time, and reaching $S_1^{(k-1)}$. $\mathcal{P}^{(k-1)}$ and $\mathcal{P'}^{(k-1)}$ together ensure that this system has a path from the canonical state $00..0$ to a state $S_{k-1}^{(k-1)}$.

Now we add delays to the edges on $\node{X}_{k}$. If we add $N-k+1$ delays on the edges from $\{\node{X}_{i\geq k}\}$, we follow $\mathcal{P'}^{(k-1)}$ and update the new delay nodes to their ON states at a suitable step of the path. When $\node{X}_{k}$ is updated to ON, the new delay nodes are still in their OFF state (because the states where nodes $\{\node{X}_{i\geq k}\}$ are ON are not visited yet), allowing this transition. Then we update the delay node on the self-edge for $\node{X}_{k}$ to ON. This does not hinder $\node{X}_{k}$ from turning OFF in order for the system to reach a memory variant of $00..0$ in the path. When the path visits a state where any of the nodes $\{\node{X}_{j>k}\}$ are ON, update the corresponding delay node to be in the ON state. Since $\node{X}_{k}$ does not have to be updated again, these delays can be kept until the system reaches $S_1^{(k)}$ (identical to $S_1^{(k-1)}$ but with the new delays in their ON states). Let's call this path $\mathcal{P'}_1^{(k)}$.

Instead, if we add $k-1$ delays on the edges from nodes $\{\node{X}_{1<i<k}\}$, we follow $\mathcal{P'}^{(k-1)}$, but now keeping the new delay nodes to their OFF states. The new delay nodes do not hinder $\node{X}_{k}$ from turning ON from $00..0$ (because the delay nodes are OFF) nor from turning OFF again (because $\bneg \node{X}_{k}$ is part of the update function of $\node{X}_{k}$). Let's call this path $\mathcal{P'}_2^{(k)}$. $\mathcal{P'}_2^{(k)}$ leads to a state $S_1^{(k)}$ that has the new delay nodes in their OFF states.

Regardless of the choice between $\mathcal{P'}_1^{(k)}$ or $\mathcal{P'}_2^{(k)}$, the system can then follow a path $\mathcal{P^{(k)}}$ almost identical to $\mathcal{P}^{(k-1)}$, leading to $S_{k-1}^{(k)}$ (identical to $S_{k-1}^{(k-1)}$ but with the new delays), as it does not involve the update of $\node{X}_{k}$. Finally, at state $S_{k-1}^{(k)}$, we turn ON $\node{X}_{k}$ to transition to $S_{k}^{(k)}$, either by using the ON states of $N-k+1$ delays coming from $\mathcal{P'}_1^{(k)}$ to use the transition from $11..1$, or by using the OFF states of $k-1$ delays coming from $\mathcal{P'}_2^{(k)}$ to use the transition from $00..0$.

\textbf{Claim 4:} Always choosing the smaller number of delays for each $\node{X}_{k}$ gives $\floor{N^2/4}$ delays.

Always choosing the smaller number of delays for any $k$ gives $\sum_{k=2}^{N} min(k-1, N-k+1)$, which is $N^2/4$ for even $N$ and $(N^2-1)/4$ for odd $N$.

\textbf{Claim 5:} It is not possible to eliminate a star-shaped MAA with fewer delays than those indicated by the algorithm. 

Suppose that the star-shaped MAA is eliminated in a system with a certain number of delay nodes. This means that there must exist a path of state transitions from the canonical $00..0$ state to the trap space $11..1$. For this path to exist, for any $0\leq k<N$, the path must contain a state $S_{k}$ (not necessarily unique) wherein $k$ non-delay nodes are in their ON states and which has a transition to a state $S_{k+1}$ where $k+1$ non-delay nodes are ON. Consider one such transition and suppose, without loss of generality, that node $\node{X}_{k+1}$ turns ON in this transition. In the absence of delays, activating $\node{X}_{k+1}$ requires all inputs to $\node{X}_{k+1}$ to be OFF or for all inputs to $\node{X}_{k+1}$ to be ON. Therefore, in $S_{k}$, the node $\node{X}_{k+1}$ must read either a memory state wherein the delay nodes copying the states of the $k$ non-delay nodes that are ON are set to OFF, or a memory state wherein the delay nodes copying the states of the $N-k$ non-delay nodes that are OFF are set to ON. In other words, the node $\node{X}_{k+1}$ must observe either all regulators as ON, or all regulators as OFF to update, but since $k$ non-delay nodes are ON, and $N-k$ non-delay nodes are OFF, we need to ``mask'' either of these groups through delays. This requires a minimum of $k$ or $N-k$ delays, respectively, meaning that any transition from any $S_{k}$ to any $S_{k+1}$ requires $\min(k, N-k)$ delay nodes on $\node{X}_{k+1}$. Therefore, at least $\sum_{k=0}^{N-1} \min(k, N-k)$ delays must exist for the whole system. 

In summary, we have shown via an algorithm that it is possible to eliminate the star-shaped MAA with $\floor{N^2/4}$ delay nodes. We have also shown that it is not possible to destroy the star-shaped MAA using fewer delay nodes. Therefore, the minimum number of delays necessary and sufficient to destroy the star MAA is $\floor{N^2/4}$.

We verified computationally that the formula above correctly indicates the required number of delays by exhaustively trying all delay combinations of size $\floor{N^2/4}-1$ on star-shaped MAAs for up to $N=5$.

\paragraph{Illustration of the algorithm and Claims 1-3 for $N=4$:}

In the following we illustrate the elements of the claims above in the case of $N=4$. 

\begin{figure}
    \centering
    \includegraphics[width=1\linewidth]{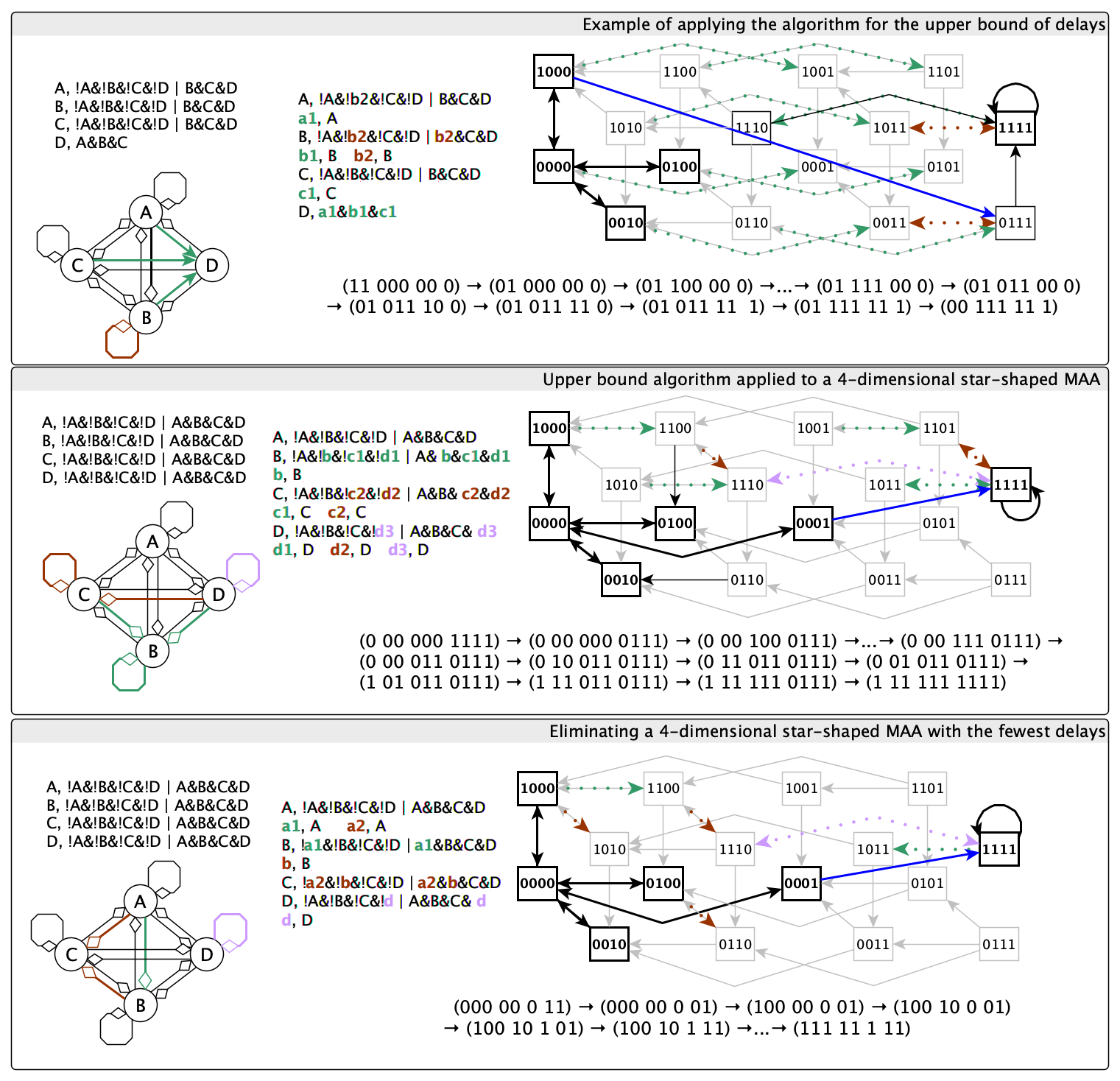}
    \caption{Illustration of the motif-avoidant attractor elimination algorithm and the algorithm that applies specifically to star-shaped MAAs as in Example \ref{ex:star_graph_family}. The visualization of interaction graphs and projected STGs is the same as in previous figures. The edges of the interaction graphs are colored by the sequence of nodes to turn ON. Top panel: A four-variable system with $d=n=3$, $m=1$.  As node $\node{D}$ has a fixed state in the MAA, the first step is to add delays to the edges from $\node{A}$, $\node{B}$, and $\node{C}$ to $\node{D}$ (green); these allow escape from the $***0$ subspace. A delay on the self-regulation of $\node{B}$ (purple) allows $\node{B}$ to turn ON in a memory version of $0011$, creating a path to the trap space. The middle and bottom panels compare the algorithm that uses the transitions of the trap space with one possible application of the specific algorithm presented in Appendix \ref{sec:supplementary_text} on a 4-dimensional star-shaped MAA. An alternative possibility for the latter is to replace the delays on the edges from $\node{A}$ to $\node{C}$ and  from $\node{B}$ to  $\node{C}$ with delays on the edge from $\node{D}$ to $\node{C}$ and the self-regulation of $\node{C}$ (see Eq. \ref{4star_variant2}).
    }
    \label{fig:star_MAA_elimination}
\end{figure}

The upper bound algorithm for $N=4$ indicates that the elimination of the MAA requires 6 delays: $\node{X}_2$ needs delays from $N-1=3$ nodes $\{\node{X}_2, \node{X}_3, \node{X}_4\}$, $\node{X}_3$ needs delays from $\{\node{X}_3, \node{X}_4\}$, and $\node{X}_4$ needs a delay on its self-regulation. The middle panel of Figure \ref{fig:star_MAA_elimination} visualizes this delay-expanded system (using an alphabetical node notation for simplicity). The figure indicates a shortest path between a canonical state derived from the original system's MAA to the trap space. The succession of states in this path is an instance of $\mathcal{P'}_1$ (the first two rows of states) followed by $\mathcal{P}$ (the last row). The first part of $\mathcal{P'}_1$ (the transitions from the canonical $00..0$ state to the canonical state with $\node{D}$ ON) are not included in this shortest path.

We illustrate the paths $\mathcal{P}$ and $\mathcal{P'}$ in the case that the first delay node to be added, $\node{d}_{1,2}$, is on the edge from from $\node{X}_1$ to $\node{X}_2$, and we are at the step of considering $\node{X}_3$ (i.e., $k=3$). We indicate the delay-expanded system's states in the order $\node{X}_1$, $\node{d}_{1,2}$, $\node{X}_2$, $\node{X}_3$, $\node{X}_4$. In this case, the path $\mathcal{P}$, which in general connects $S_0$ to $S_{k-1}$ (the state before updating $\node{X}_k$), is $\mathcal{P}^{(2)}$: $(00~0~0~0)\rightarrow (10~0~0~0) \rightarrow (10~1~0~0)$.  The generic $\mathcal{P'}$ path starts at the canonical state $00..0$ in the MAA, visits states where each of the nodes is ON individually, going back to a canonical or memory variant of $00..0$ each time, and reaches a state $S_1$ (which has $\node{X}_1=1$). In this case, $\mathcal{P'}$ is $\mathcal{P'}_2^{(2)}$: $(00~0~0~0) \rightarrow (10~0~0~0) \rightarrow (00~0~0~0) \rightarrow (00~1~0~0) \rightarrow (00~0~0~0) \rightarrow (00~0~1~0) \rightarrow (00~0~0~0) \rightarrow (00~0~0~1) \rightarrow (00~0~0~0) \rightarrow (10~0~0~0)$.

Next, there are two choices of delay combinations to add to $\node{X}_3$: from $\node{X}_1$ and $\node{X}_2$ (as in Claim 2), or on the self-regulation of $\node{X}_3$ and from $\node{X}_4$ (as in Claim 1). The delay-expanded system for the first choice is indicated in  Equation \ref{4star_variant1}, and the delay-expanded system for the second choice is in Equation \ref{4star_variant2}. 

\begin{align}
\label{4star_variant1}
\node{X}_1,&~(\bneg \node{X}_1 \band \bneg \node{X}_2 \band \bneg \node{X}_3 \band \bneg \node{X}_4) \bor (\node{X}_1 \band \node{X}_2 \band \node{X}_3 \band \node{X}_4) \nonumber\\
\node{d}_{1,2},&~\node{X}_1 \nonumber\\
\node{d}_{1,3},&~\node{X}_1 \nonumber\\
\node{X}_2,&~(\bneg\node{d}_{1,2} \band \bneg \node{X}_2 \band \bneg \node{X}_3 \band \bneg \node{X}_4) \bor (\node{d}_{1,2} \band \node{X}_2 \band \node{X}_3 \band \node{X}_4) \nonumber\\
\node{d}_{2,3},&~\node{X}_2 \nonumber\\
\node{X}_3,&~(\bneg \node{d}_{1,3} \band \bneg \node{d}_{2,3} \band \bneg \node{X}_3 \band \bneg \node{X}_4) \bor (\node{d}_{1,3} \band \node{d}_{2,3} \band \node{X}_3 \band \node{X}_4) \nonumber\\
\node{X}_4,&~(\bneg \node{X}_1 \band \bneg \node{X}_2 \band \bneg \node{X}_3 \band \bneg \node{d}_4) \bor (\node{X}_1 \band \node{X}_2 \band \node{X}_3 \band \node{d}_4) \nonumber\\
\node{d}_4,&~\node{X}_4
\end{align}

The path $\mathcal{P'}_2^{(3)}$ is a specific instance of the path $\mathcal{P'}$ applied to the case when delay nodes are added from $\node{X}_1$ and $\node{X}_2$, and the state of these delay nodes is OFF. The system state is now given in the order $\node{X}_1$, $\node{d}_{1,2}$, $\node{d}_{1,3}$, $\node{X}_2$, $\node{d}_{2,3}$, $\node{X}_3$, $\node{X}_4$. In this case $\mathcal{P'}_2^{(3)}$ is $(000~00~0~0)\rightarrow (100~00~0~0) \rightarrow (000~00~0~0) \rightarrow (000~10~0~0) \rightarrow (000~00~0~0) \rightarrow (000~00~1~0) \rightarrow (000~00~0~0) \rightarrow (000~00~0~1) \rightarrow (000~00~0~0) \rightarrow (100~00~0~0)$. 

\begin{align}
\label{4star_variant2}
\node{X}_1,&~(\bneg \node{X}_1 \band \bneg \node{X}_2 \band \bneg \node{X}_3 \band \bneg \node{X}_4) \bor (\node{X}_1 \band \node{X}_2 \band \node{X}_3 \band \node{X}_4) \nonumber\\
\node{d}_{1,2},&~\node{X}_1 \nonumber\\
\node{X}_2,&~(\bneg \node{d}_{1,2} \band \bneg \node{X}_2 \band \bneg \node{X}_3 \band \bneg \node{X}_4) \bor (\node{d}_{1,2} \band \node{X}_2 \band \node{X}_3 \band \node{X}_4) \nonumber\\
\node{X}_3,&~(\bneg \node{X}_1 \band \bneg \node{X}_2 \band \band \node{d}_3 \band \bneg \node{d}_{4,3}) \bor (\node{X}_1 \band \node{X}_2 \band \node{d}_3 \band \node{d}_{4,3}) \nonumber\\
\node{d}_3,&~\node{X}_3 \nonumber\\
\node{X}_4,&~(\bneg \node{X}_1 \band \bneg \node{X}_2 \band \bneg \node{X}_3 \band \bneg \node{d}_4) \bor (\node{X}_1 \band \node{X}_2 \band \node{X}_3 \band \node{d}_4) \nonumber\\
\node{d}_{4,3},&~\node{X}_4 \nonumber\\
\node{d}_4,&~\node{X}_4
\end{align}

The path $\mathcal{P'}_1^{(3)}$ is a specific instance of the path $\mathcal{P'}$ applied to the case when delay nodes are added on the self-regulation of $\node{X}_3$ and from $\node{X}_4$, and the state of these delay nodes is ON. The system state is now given in the order $\node{X}_1$, $\node{d}_{1,2}$, $\node{X}_2$, $\node{X}_3$, $\node{d}_3$, $\node{X}_4$, $\node{d}_{4,3}$. In this case $\mathcal{P'}_1^{(3)}$ is $(00~0~00~00) \rightarrow (10~0~00~00) \rightarrow (00~0~00~00) \rightarrow (00~1~00~00) \rightarrow (00~0~00~00) \rightarrow (00~0~10~00) \rightarrow  (00~0~11~00) \rightarrow (00~0~01~00) \rightarrow (00~0~01~10) \rightarrow (00~0~01~11) \rightarrow (00~0~01~01) \rightarrow  (10~0~01~01)$. This path involves the canonical $00..0$ state until the delay node on the self-regulation of $\node{X}_3$ is updated and involves memory states afterwards.

The bottom panel of Figure \ref{fig:star_MAA_elimination} indicates an example of a shortest path from a canonical state derived from the original system's MAA to the trap space $11..1$ in the case of the system of Equation \ref{4star_variant1}. This path starts with the minimal necessary step of $\mathcal{P'}_2^{(4)}$, namely the start from the canonical state with $D=1$, then turning $\node{D}$ OFF, it is followed by $\mathcal{P}^{(4)}$, namely the succession of memory states that turn $\node{A}$, then $\node{B}$, then $\node{C}$, then $\node{D}$ ON, and ends with the delay nodes turning ON as well.

In a confirmation of the algorithm and the claims, we found that the $N=4$ systems with the minimal combinations of delays that no longer have an MAA are $N!=24$ permutations of Equation \ref{4star_variant1} and $24$ permutations of Equation \ref{4star_variant2}. These combinations of delays are the two possibilities of minimal delay combinations indicated by the algorithm. For $N=5$ we verified that no combinations of five delays can eliminate the MAA. We also verified that the 6-delay combination indicated by the algorithm, with delays on the edges from $\node{X}_1$ to $\node{X}_2$, from $\node{X}_1$ to $\node{X}_3$, from $\node{X}_2$ to $\node{X}_3$, the self-edge on $\node{X}_4$, the edge from $\node{X}_5$ to $\node{X}_4$, and the self-edge of $\node{X}_5$, successfully eliminates the MAA.

\subsection{Not all memory states are reachable from canonical states}
\label{sec:unreachability_issue}

As we described in the main text, the canonical states of a delay-expanded system retain the reachability of the original system. Yet, a non-canonical (memory) state may gain a transition compared to the original system (i.e., the delay-affected node changes state although its state stays the same in the original system) or lose a transition (i.e. the delay-affected node no longer changes state).

Lost transitions of memory states can interfere with the capacity of gained transitions to eliminate the MAA. For example, even if a memory state gains a transition that connects it to the trap space, this memory state may not be reachable from the canonical states of the motif-avoidant attractor because another memory state has lost a transition. In these cases of unreachability the motif-avoidant attractor is preserved. We identified that this unreachability is not avoidable in general, except for memory states due to the linear extension of a self-edge. In this case, the canonical states have transitions to memory states whenever the delay-affected node (which is also the delay node's parent node) is updated. Since the connectivity of canonical states is preserved, these memory states take part in the strongly connected component of the state transition graph as the mediators of paths between canonical states. When the nodes unaffected by delays are updated, all memory states retain the original connectivity. This ensures that all memory states of the MAA are reachable when self-edges are extended.

\begin{figure}
    \centering
    \includegraphics[width=1\linewidth]{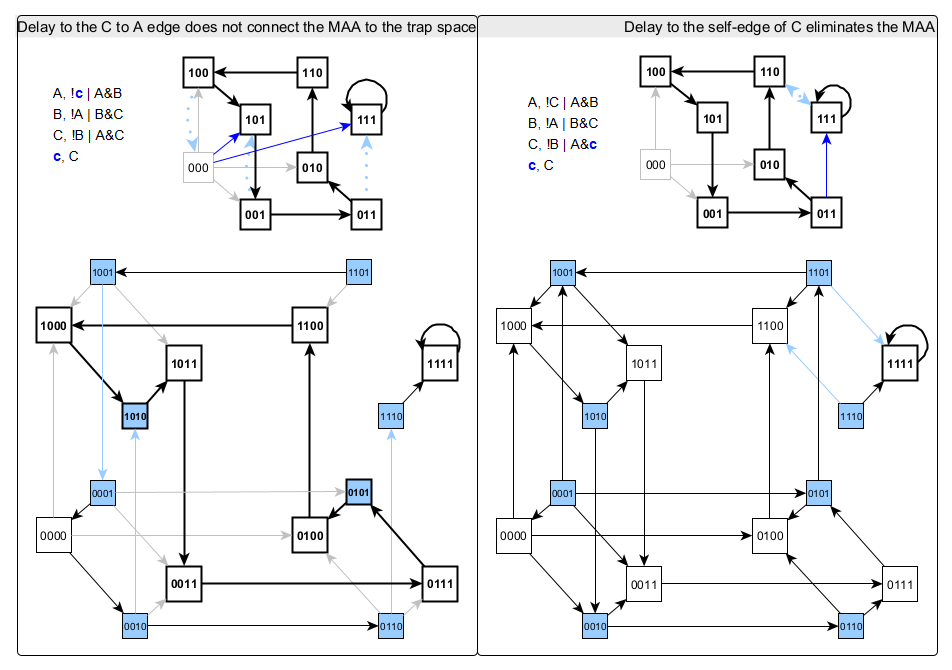}
    \caption{ Illustration of the unreachability of certain memory states from canonical states. We expand the system in Example \ref{ex:template2iv} with a delay to the inhibition of $\node{A}$ by $\node{C}$ (left panel) or a delay to the self-regulation of $\node{C}$. For both cases we indicate the full state transition graph as well as the projected state transition graph. In the state transition graph of the delayed systems the canonical states have white background and the memory states have light blue background. The state transitions that are different for memory states compared to canonical states are marked in light blue in both the full and projected state transition graphs. In each panel the strongly connected components of states corresponding to the original (delay-free) system's motif-avoidant attractor is indicated  with thick edges and thick node outlines. Each dark blue edge of the projected state transition graph arises from a path among canonical states in the delayed system. }
    \label{fig:not_all_delays_work}
\end{figure}

An example of a failed MAA elimination due to an unreachable memory state is offered by the system in Example \ref{ex:template2iv}, and is illustrated in Figure \ref{fig:not_all_delays_work}. The MAA visits all three states that are at Hamming distance 1 from the trap space $111$, thus turning a single state ON from a state in which the other two are already ON should be a viable strategy to eliminate the MAA. A possible way to turn node $\node{A}$ ON in a memory version of the $011$ state is to add a delay on the $\node{C}$ to $\node{A}$ edge. This edge is not part of the regulations that maintain the trap space. Such a delay allows $\node{A}$ to turn ON from the memory state $0110$ (see the light blue edge) and creates a path to the trap space. However, the memory state 0110 is not accessible from the strongly connected component of the state transition graph because the memory version of the state $101$ lost its transition to the memory version of 001 which existed in the original system (compare the light blue edge among memory states to the bold black edge among canonical states). As a consequence, the motif-avoidant attractor is preserved. Another way to turn node $\node{A}$ ON in a memory version of the $011$ state is to add a delay to the self-regulation of $\node{A}$, which is part of the stable motif, and use the memory of a past ON state of $\node{A}$. The projected state transition graph for that delay is indicated in Figure \ref{fig:3_prototypical_MAAs}. Here we indicate the equivalent scenario of adding a delay to the self-edge of $\node{C}$, which also is part of the regulations that maintain the trap space. The delay leads to two new state transitions in the delayed system, including a transition from the memory state $1101$ to the trap space $1111$. The memory state $1101$ is part of the strongly connected component of states in the delayed system, thus the new transition places the trap space in the out-component of the strongly connected component. As the out-component is not empty, there no longer is a motif-avoidant attractor. In summary, the unreachability issue makes a delay on the $\node{C}$ to $\node{A}$ edge unable to eliminate the MAA, but delays on the self-regulation of any of the three nodes is able to eliminate the MAA.

Notably, the example illustrates that memory states lose transitions only when the delay-affected node is updated. Our algorithm for the general upper bound and the specific solution for the star-shaped MAA avoid the unreachability issue by identifying a path that allows all necessary delay nodes to be updated to the needed states \emph{after} the delay-affected node is updated to its state that is opposite of the trap space. The delay nodes can retain these states until the delay-affected node is updated again, and allow it to align with the trap space.

At present there is no estimate of the likelihood of a memory state needed to create a path to the trap space being unreachable from the motif-avoidant attractor. We observed that unreachable memory states are fairly common, but this unreachability preventing a path to the trap space occurs rarely. We have tried and failed to find examples of unsuccessful MAA elimination due to unreachability in cases where $m>1$. (Recall that $m<d$ represents the minimal disagreement between a state of the MAA with the fixed nodes of the trap space.) In all of our examples, we observed that multiple combinations of delays could create paths to the trap space, and only a subset of these paths were affected by unreachability issues. At least one path remained that did successfully eliminate the MAA.

If in fact it is true that unreachability only occurs for $m=1$, then this issue does not pose a problem to MAA elimination. In the case of $m=1$, the MAA can be eliminated by adding a delay node to a self-regulation. For different values of $m$, delays on edges within the MAA could be used to lower the number of delays needed to eliminate the MAA.

\section{Supplementary Figures}

\begin{figure}
    \centering
    \includegraphics[width=.7\linewidth]{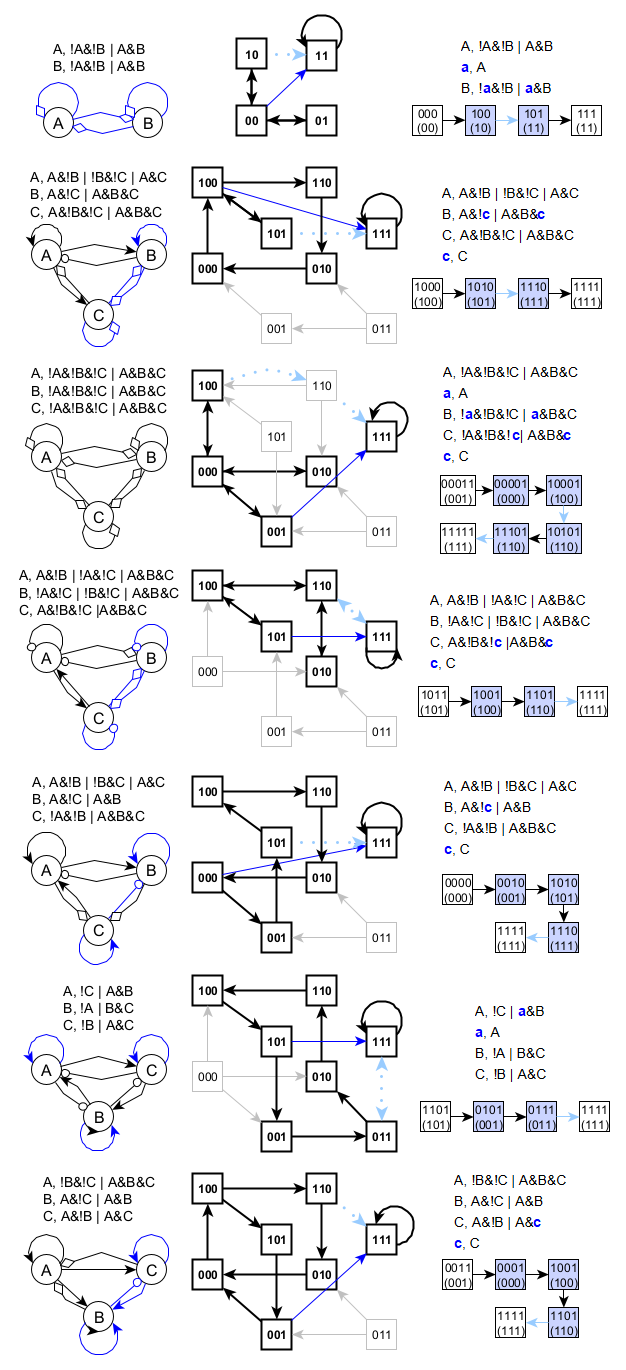}
    \caption{Prototypical (template) MAAs with minimal connectivity in the state transition graph (i.e. eliminating any state transition would eliminate the MAA). For each MAA we indicate the interaction graph, set of Boolean functions, and projected STGs of a system that also has a point attractor $111$. All but one MAA have single delays that eliminate the MAA; the edges whose linear extension eliminates the MAA are shown in blue in the interaction graph.  For each MAA we indicate one example of a single (or in one case, double) delay that eliminates the MAA. In the projected STGs the states and state transitions that make up attractors are shown in bold. The state transitions due to the delay nodes (and involving memory states) are shown with dotted light blue edges. The blue edges among canonical states arise from paths of state transition in the delayed system; these paths are indicated below the projected STG and highlight the state of each delay node in blue.  }
    \label{fig:prototypical_MAAs}
\end{figure}

\begin{figure}
    \centering
    \includegraphics[width=1\linewidth]{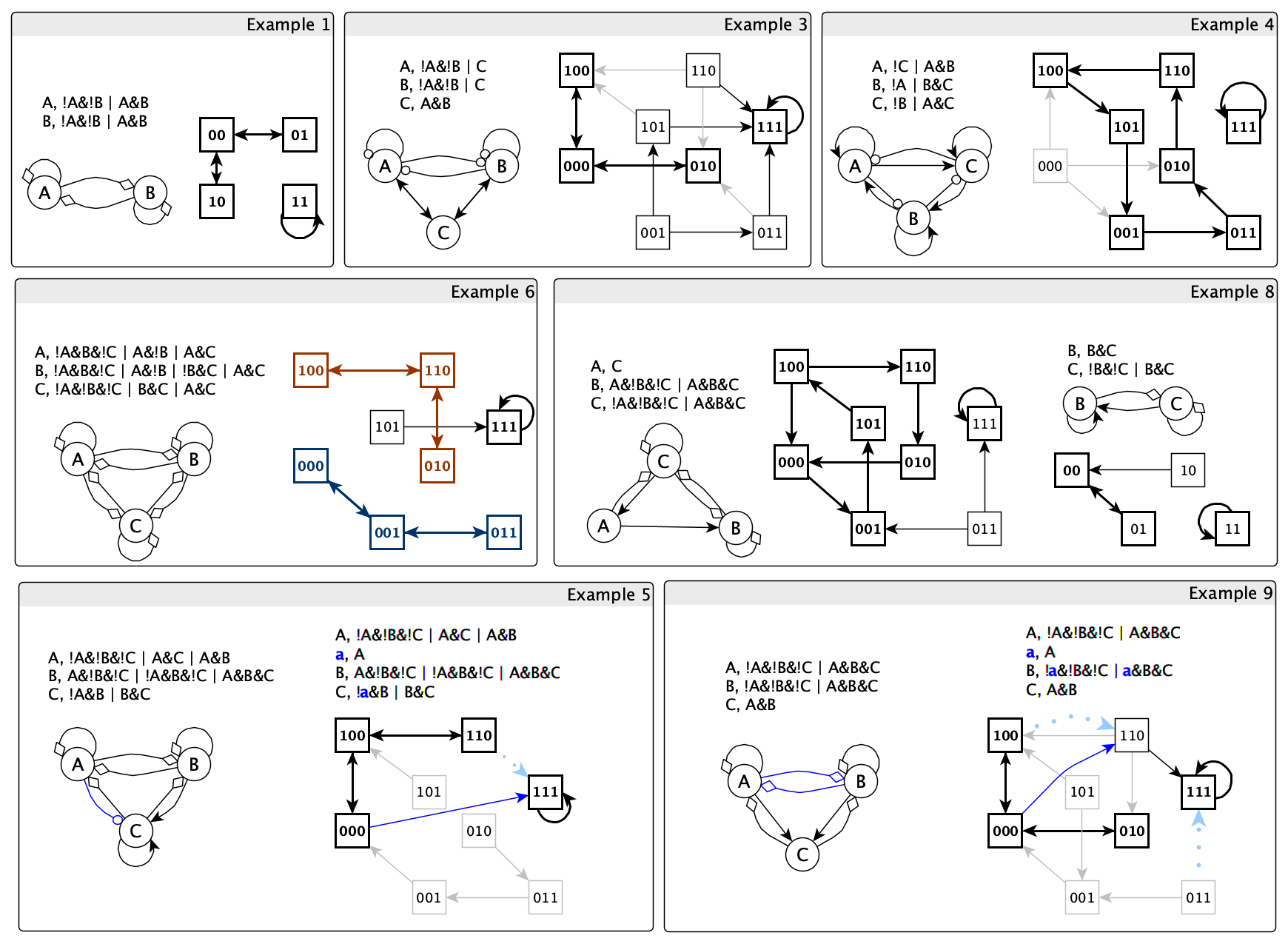}
    \caption{The interaction graphs and state transition graphs of the systems in Examples 1, 3-6, 8 and 9. In the interaction graph a terminal arrow indicates a positive edge, a terminal circle indicates a negative edge and a terminal diamond indicates a dual edge. In the state transition graph the attractors are indicated in bold outlines. The interaction graphs of Examples \ref{ex:out_and_in_transition} and \ref{ex:two_single_delays} indicate in blue the edges whose linear extension eliminates the motif-avoidant attractor. We indicate the projected state transition graph corresponding to one linear extension.  Example \ref{ex:out_and_in_transition} illustrates that escaping the subspace of the MAA can be accomplished by linear extension of a subset of the edges incident on the nodes fixed by the MAA. Example \ref{ex:two_single_delays} illustrates that when the state that escapes the subspace can reach the trap space, it is enough to use delays that allow reaching this escape state, and no delays to the fixed node(s) are needed.}
    \label{fig:example_figures}
\end{figure}

\begin{figure}
\centering
\begin{subfigure}{0.6\linewidth}
\small
\definecolor{custom_green}{HTML}{228833}
\setlength\tabcolsep{2pt} 
\begin{tabular}{| c | c | c | c | c | c | c | c |}
    \hline
    K\textbackslash N & 4 & 5 & 6 & 7 & 8 & 9 & 10 \\\hline
     2 &  \cellcolor{custom_green!6} \texttt{1.7\%} & \cellcolor{custom_green!11} \texttt{2.9\%} & \cellcolor{custom_green!17} \texttt{4.4\%} & \cellcolor{custom_green!22} \texttt{5.8\%} & \cellcolor{custom_green!32} \texttt{8.2\%} & \cellcolor{custom_green!38} \texttt{9.8\%} & \cellcolor{custom_green!50} \texttt{12.8\%} \\\hline
     3 & \cellcolor{custom_green!4} \texttt{1.0\%} & \cellcolor{custom_green!6} \texttt{1.6\%} & \cellcolor{custom_green!9} \texttt{2.4\%} & \cellcolor{custom_green!13} \texttt{3.4\%} & \cellcolor{custom_green!17} \texttt{4.6\%} & \cellcolor{custom_green!20} \texttt{5.3\%} & \cellcolor{custom_green!27} \texttt{6.9\%} \\\hline
     4 & \cellcolor{custom_green!2} \texttt{0.5\%} & \cellcolor{custom_green!3} \texttt{0.9\%} & \cellcolor{custom_green!5} \texttt{1.4\%} & \cellcolor{custom_green!7} \texttt{1.8\%} & \cellcolor{custom_green!9} \texttt{2.4\%} & \cellcolor{custom_green!11} \texttt{3.0\%} & \cellcolor{custom_green!14} \texttt{3.8\%} \\\hline
     5 & - & \cellcolor{custom_green!1} \texttt{0.4\%} & \cellcolor{custom_green!2} \texttt{0.6\%} & \cellcolor{custom_green!3} \texttt{0.9\%} & \cellcolor{custom_green!4} \texttt{1.1\%} & \cellcolor{custom_green!6} \texttt{1.5\%} & \cellcolor{custom_green!7} \texttt{1.9\%} \\\hline
     6 & - & - & \cellcolor{custom_green!1} \texttt{0.4\%} & \cellcolor{custom_green!2} \texttt{0.5\%} & \cellcolor{custom_green!2} \texttt{0.6\%} & \cellcolor{custom_green!3} \texttt{0.8\%} & \cellcolor{custom_green!3} \texttt{0.9\%} \\\hline
     7 & - & - & - & \cellcolor{custom_green!1} \texttt{0.3\%} & \cellcolor{custom_green!1} \texttt{0.4\%} & \cellcolor{custom_green!2} \texttt{0.5\%} & \cellcolor{custom_green!2} \texttt{0.6\%} \\\hline
     8 &- & - & - & - & \cellcolor{custom_green!1} \texttt{0.3\%} & \cellcolor{custom_green!1} \texttt{0.3\%} & \cellcolor{custom_green!1} \texttt{0.4\%} \\\hline
     9 & - & - & - & - & - & \cellcolor{custom_green!1} \texttt{0.3\%} & \cellcolor{custom_green!1} \texttt{0.3\%} \\\hline
     10 & - & - & - & - & - & - & \cellcolor{custom_green!1} \texttt{0.2\%} \\\hline
     \multicolumn{8}{|c|}{21 301 660 total networks} \\\hline
\end{tabular}
\caption{Small networks}
\end{subfigure}
\small
\definecolor{custom_green}{HTML}{228833}
\setlength\tabcolsep{2pt} 
\begin{subfigure}{0.35\linewidth}
\centering
\begin{tabular}{| c | c | c | c | }
    \hline
    K\textbackslash N & 20 & 30 & 40 \\\hline
     2 & \cellcolor{custom_green!7} \texttt{7.0\%} & \cellcolor{custom_green!16} \texttt{16.0\%} & \cellcolor{custom_green!50} \texttt{49.1\%} \\\hline
     3 & \cellcolor{custom_green!3} \texttt{3.0\%} & \cellcolor{custom_green!6} \texttt{6.2\%} & \cellcolor{custom_green!10} \texttt{10.2\%} \\\hline
     4 & \cellcolor{custom_green!1} \texttt{1.6\%} & \cellcolor{custom_green!3} \texttt{2.9\%} & \cellcolor{custom_green!1} \texttt{1.2\%} \\\hline
     5 & \cellcolor{custom_green!0} \texttt{0.8\%} & \cellcolor{custom_green!1} \texttt{1.5\%} & \cellcolor{custom_green!0} \texttt{0.02\%} \\\hline
     \multicolumn{4}{|c|}{\makecell{158 948 231\\total networks}} \\\hline
\end{tabular}
\caption{Large networks}
\end{subfigure}
    \caption{The distribution of sampled random networks across N-K ensembles. For the small networks, each ensemble was sampled until 1000 networks with motif-avoidant attractors were detected. The same criterion was implemented for the large networks, but the [N=40, K=4] and [N=40, K=5] ensembles had to be terminated prematurely, because not enough motif-avoidant attractors were discovered, even after several weeks of runtime.}
    \label{fig:random-ensemble-distribution}
\end{figure}

\end{appendices}

\end{document}